\documentclass{ieeeaccess}

\usepackage{cite}
\usepackage[english]{babel}
\usepackage{algorithm}
\usepackage[utf8]{inputenc}
\usepackage[noend]{algpseudocode}
\usepackage{amsmath,amssymb,amsfonts}
\usepackage{verbatim}
\usepackage{amsmath}
\usepackage{amsmath,amssymb,amsthm}
\usepackage{array,booktabs}
\usepackage{multirow}
\usepackage{graphicx}
\usepackage{textcomp}
\usepackage{caption}
\usepackage{comment}
\usepackage{multirow}
\usepackage{color, colortbl}
\usepackage{subfigure}
\usepackage{hyperref}
\usepackage{float}
\usepackage{nomencl}
\usepackage{wasysym}
\usepackage{soul}

\begin{document}
\history{Date of publication xxxx 00, 0000, date of current version xxxx 00, 0000.}
\doi{10.1109/ACCESS.2017.DOI}

\title{On the Feasibility of Load-Changing Attacks in Power Systems during the COVID-19 Pandemic}

\author{\uppercase{
\uppercase{Juan Ospina}\authorrefmark{1}, \IEEEmembership{Member, IEEE},
\uppercase{Xiaorui Liu}\authorrefmark{1}, \IEEEmembership{Student Member, IEEE} \\ 
\uppercase{Charalambos Konstantinou}\authorrefmark{1}, \IEEEmembership{Senior~Member, IEEE}, and  \uppercase{Yury Dvorkin}\authorrefmark{2}, \IEEEmembership{Member, IEEE}}
\address[1]{Center for Advanced Power Systems, FAMU-FSU College of Engineering, Florida State University, , Tallahassee, FL 32310, USA}\address[2]{Department of Electrical and Computer Engineering, New York University, Brooklyn, NY 11201, USA}}

\markboth
{Ospina \headeretal: On the Feasibility of Load-Changing Attacks in Power Systems during the COVID-19 Pandemic}
{Ospina \headeretal: On the Feasibility of Load-Changing Attacks in  Power Systems during the COVID-19 Pandemic}
       
\corresp{Corresponding author: Charalambos Konstantinou (e-mail: ckonstantinou@ieee.org).}

\begin{abstract}

The electric power grid is a complex cyberphysical energy system (CPES) in which information and communication technologies (ICT) are integrated into the operations and services of the power grid infrastructure. The growing number of Internet-of-things (IoT) high-wattage appliances, such as air conditioners and electric vehicles, being connected to the power grid, together with the high dependence of ICT and control interfaces, make CPES vulnerable to high-impact, low-probability load-changing cyberattacks. Moreover, the side-effects of the COVID-19 pandemic demonstrate a modification of electricity consumption patterns with utilities experiencing significant net-load and peak reductions. These unusual sustained low load demand conditions could be leveraged by adversaries to cause frequency instabilities in CPES by compromising hundreds of thousands of IoT-connected high-wattage loads. This paper presents a feasibility study of the impacts of load-changing attacks on CPES during the low loading conditions caused by the lockdown measures implemented during the COVID-19 pandemic. The load demand reductions caused by the lockdown measures are analyzed using dynamic mode decomposition (DMD), focusing on the March-to-July 2020 period and the New York region as the most impacted time period and location in terms of load reduction due to the lockdowns being in full execution. Our feasibility study evaluates load-changing attack scenarios using real load consumption data from the New York Independent System Operator (NYISO) and shows that an attacker with sufficient knowledge and resources could be capable of producing frequency stability problems, with frequency excursions going up to 60.5 Hz and 63.4 Hz, when no mitigation measures are taken.
\end{abstract}



\titlepgskip=-15pt

\maketitle

\section{Introduction}
The wide-scale deployment of information, sensing, and communication technologies in electric power systems (EPS) contribute to various power grid functionalities. For example, Internet-of-Things (IoT) devices are widely utilized in industrial assets to provide control and monitoring support for supervisory control and data acquisition (SCADA), advanced metering infrastructure (AMI), and other types of communication and control infrastructures. 
As a result, during the last few years, the efficiency, robustness, and reliability of cyberphysical energy systems (CPES) have been greatly improved. 
At the same time, the vulnerabilities inherited from the IoT ecosystem have expanded the CPES threat surface. The roll-out of customer-end IoT-controllable high-wattage devices and distributed energy resources (DERs) unlocks new vulnerabilities on the demand-side of CPES and opens new avenues for adversaries to launch large-scale coordinated remote attacks on power system's assets. For example, DERs or modern controllable loads use IoT devices to coordinate their operations with other CPES (e.g., via  IoT-based smart meters or other home assistants such as Amazon Echo and Google Home). Naturally, these IoT interfaces become great attack vectors due to the countless vulnerabilities engendered by the complex IoT supply chains. Insecure remote login passwords of IoT devices can be exploited for malware infection as in the Mirai botnet which, in 2016, successfully compromised thousands of household IoT devices in a distributed denial-of-service campaign \cite{antonakakis2017understanding}. Firmware updates are another source of contamination \cite{konstantinou2016taxonomy}, making consumer devices such as printers \cite{Cui2013WhenFM} and relay controllers \cite{konstantinou2015impact, wang2015confirm} vulnerable to firmware attacks. Ghost domain name system (DNS) attacks can infect IoT devices by routing them to malicious DNS servers  \cite{Jiang2012GhostDN}. Attacks on electricity consumers and DERs are essentially facilitated by the negligence of the customers that use default or weak security credentials \cite{wired_2018}.

Adversaries capable of compromising IoT-controllable high wattage loads and DERs of CPES can maliciously affect the stability of the grid, causing degradation of grid equipment or even power outages and large-scale blackouts \cite{soltan2018blackiot}.  Although currently hypothetical due to the low penetration rates of high-wattage loads and DERs (prosumers), these types of attacks are projected to become realistic in the near future as the penetration rates are anticipated to grow exponentially \cite{cyberassessment}. Due to the distributed nature of \emph{load-changing attacks}, which disturb the system from the demand-side, as well as the various ways of attack payload weaponization \cite{raman2020weaponizing}, it is difficult for system operators to detect and mitigate such attacks in order to maintain frequency stability and ensure that voltage and frequency levels are within operational limits. In addition, IoT-based attacks to CPES do not require operational knowledge of the power system and are very easy to repeat. 


Starting in early 2020, the COVID-19 pandemic has directly affected businesses and individuals, significantly increasing their dependency on the Internet (e.g., remote work). This reliance expanded the number of cyberspace-related incidents as malicious attempts have proliferated to exploit this sudden unplanned shift in society. COVID-19 drives criminal and political cyberattacks across networks, cloud, and mobile phones; pandemic-related attacks increased exponentially from under 5,000 per week in February 2020, to over 200,000 per week in late April 2020 \cite{cyberattackreport}. For instance, COVID-19 scams often use phishing emails and malicious websites to promote fake vaccines and cures, fraudulent charity drives, and false information on government aid, while at the same time delivering malware to unsuspecting users and insecure services. Table \ref{table:covid19threats} shows a summary of some of the most prominent threats and vulnerabilities exacerbated due to the COVID-19 pandemic and lockdown measures. Example attack categories range from malware, Zoom bombing, and phishing attacks all the way to sophisticated critical infrastructure attacks that could compromise the integrity of the power grid. Due to the barrage of cyberattacks during COVID-19, the malicious exploitation potential of ICT and control interfaces of IoT-controllable loads could, more than ever before, constitute a realistic cyber-threat to power grid operations.

\begin{table}[t]
\centering
\caption{Threats and vulnerabilities increased during COVID-19 pandemic.}
\label{table:covid19threats}
\begin{tabular}{||c|c||}
\hline\hline
\textbf{Threats and vulnerabilities} & \textbf{Online resource} \\ \hline
Malware and phishing & \cite{malw1} \cite{malw2} 
\\ \hline
Spyware and phishing & \cite{spw1} 
\\ \hline
Ransomware & \cite{spw1}
\\ \hline
ZOOM bombing & \cite{zb1} \cite{zb2} 
\\ \hline
Health check -- ISPs, cloud providers, &\multirow{2}{*}{\cite{isp}
}\\
UCaaS during pandemic & \\ \hline
Critical power grid infrastructure attacks & \cite{crit1} \cite{crit2} \\ \hline \hline
\end{tabular}
\end{table}

\vspace{6mm}

\section{Related Work \& Contributions} 

\subsection{Load-Changing Attacks}
Existing work on load-changing attacks\footnote{Prior work has used the names ``manipulation of demand
via IoT attack (MadIoT)'' \cite{soltan2018blackiot}, ``dynamic load-altering attack (DLAA)'' \cite{amini2016dynamic}, ``coordinated load changing attack'' \cite{dabrowski2017grid},  ``demand-side cyberattacks'' \cite{acharya2020public}, and ``power botnet attacks'' \cite{wang2019securing}. Our terminology of \emph{load-changing attacks} refers to such attack strategies aiming in a coordinated way to control IoT connected high-wattage appliances and DERs.} has focused on the modeling and analysis of attack vectors in which system conditions, e.g., electric power demand and electricity prices, make use of forecasting models which do not account for sudden reductions in peak demand and delivered energy nor modification of consumption patterns \cite{paaso2020sharing}. The authors in \cite{huang2019not} have  demonstrated that load-changing attacks can cause controlled load shedding but not cascading failures. Sudden IoT attacks increasing the system load demand will cause under frequency load shedding (UFLS) control to split the frequencies of the buses into islands of different operating regions of the grid when the system cannot handle the load and the frequency drops. In \cite{acharya2020public}, the authors investigated the feasibility of a load-changing attack using compromised electric vehicles (EVs). After canvassing publicly available data, they recovered the exact configuration of the high-voltage power grid and public EV charging stations (EVCS) in Manhattan, NY. Using this publicly available data, the work designed a data-driven attack mechanism requiring from around 500 to about 5,000 compromised Tesla EVs, depending on grid conditions and attack parameters, to destabilize the frequency, leading to a major blackout. Furthermore, a dynamic load-changing attack against power system stability is studied in \cite{amini2016dynamic}. The authors formulated the problem as a close-loop attack in which a dilettante adversary controls the changes in the compromised load based on the system frequency feedback.

\subsection{Impact of COVID-19 in Power Systems}

The recent novel coronavirus disease (COVID-19) severely affected the entire globe as a public health and economic crisis. At the same time, the COVID-19 pandemic caused substantial changes in the operations of bulk power systems and electricity markets. One of the most evident effects of the health crisis on EPS is the reduction in peak demand and consumed energy; electricity demand remains lower than typical expectations during the pandemic in many regions of the world. For example, the electricity demand in June 2020 in EU countries was 10\% below the 2019 levels. In Italy, the 2019-2020 year-on-year change in electricity demand, weather corrected, reached 28\% during the 15th week of 2020 \cite{Covid-19ioe}. Similarly in Spain, the electricity consumption decreased by 13.5\% during the March-to-April period when compared to the average load consumption of the last five years due to the lockdown measures implemented to stop the spread of the COVID-19 pandemic \cite{santiago2020electricity}. The reduction was mainly observed during working days with an average 14.5\% reduction while during weekends the reduction observed was around 10.6\%. In the African region, one example is the study conducted by researchers in \cite{fallingconsumption}, where an exploratory research aimed at investigating the impacts caused by the COVID-19 pandemic in the South African power grid was performed. According to the authors, the electricity consumption and peak demand in South Africa decreased by 28.1\% and 20.2\%, respectively; exacerbating issues in the reliability of the South African electric grid. Furthermore, in the East Asian region, and more specifically in China, a 13\% electricity demand drop, during the lockdown period, was observed  \cite{Covid-19ru}.

COVID-19 lockdown measures are also being analyzed for several cities and regions of the United States (U.S.). One example is the study conducted in \cite{Covid-19ru}, where researchers found that New York City's (NYC) hourly demand in mid-April 2020 ranged from roughly 5\% to 21\% below typical levels while reductions in electric consumption averaged 21\% below expected values during the 8 am hour. Similarly, researchers in \cite{ERYILMAZ2020106829} explored the impacts that the stay-at-home orders (SAHO), issued in response to the COVID-19 pandemic, had on the regional electricity generation fuel mixes in three major Regional Transmission Operators (RTOs) in the U.S., i.e., New York Independent System Operator (NYISO), Midcontinent Independent System Operator (MISO), and Pennsylvania Jersey Maryland  (PJM). The authors use of the Shannon-Wiener diversity index, an index designed to measure the fuel diversity of a generation portfolio, in order to compare the before and after SAHO periods in terms of electricity generation fuel mixes. Even though they did not find any significant difference in the average diversity index between the two  periods, the variance (i.e., $\sigma^2$) of the index for MISO, NYISO, and PJM  was significant, with a lower 35\% value for the first two RTOs and an 8\% higher value for the latter.

The impact on energy consumption and peak demand as well as the alternation of the consumption patterns has brought numerous challenges for utilities and system operators. Different practices have been followed to proactively mitigate technical issues and maintain normal operating conditions of CPES \cite{paaso2020sharing}. Examples include switching off and on automatic voltage control systems, capacitors, and reactors, utilizing STATCOM (static synchronous compensator) and UPFC (unified power flow controller) devices to absorb reactive power, or the use of automatic UFLS and reserve arrangements, among others. In \cite{implicationsofcovid}, a comprehensive review of the implications of the COVID-19 pandemic for the electric industry is presented. In this review, the impacts that the pandemic caused in the power balance and electricity prices are explored, with the main focus on exploring how the uncertainty of load demand, voltage violation conditions, and other challenges have posed higher pressure on system operators and system maintenance. According to the authors, in most countries that applied lockdown measures, the total load demand decreased, and this, in turn, increased the uncertainty of load and posed higher requirements for load forecasting accuracy and system reserves. A similar study is presented in \cite{impactofcovid}, where authors explore the impacts of the pandemic on the U.S. electricity demand and supply. The work performs an analysis of the electricity consumption data up to the end of May 2020 and examines some of the variables (e.g., daily peak, demand ramp rate, demand forecast error, and net electricity interchange, etc.) that could indicate stress on the power grid. The study conducted is limited to three states: California, Florida, and New York, and concluded that the effect of the pandemic may not only be a simple reduction in the load, but there exist noticeable differences among the examined regions during the SAHOs. Focusing on smaller electric grids, the research in \cite{readiness} presents a detailed study regarding the impact that the COVID-19 pandemic had on the operation of the Estonian, Israeli, and Finnish grids with a special focus on three major effects: changing patterns in generation and consumption, frequency stability, and high integration of renewables. The authors concluded that the reduction of load consumption during a pandemic-type event affects considerably the management and control of generation units and leads to voltage and frequency deviations that reduce the reliability of the system. For these reasons, the addition of tools and regulation devices designed to manage abnormal events and mitigate reliability problems is suggested in the long-term planning of the grid infrastructure.

\begin{figure*}[t]
\centering
  \includegraphics[width=0.90\textwidth]{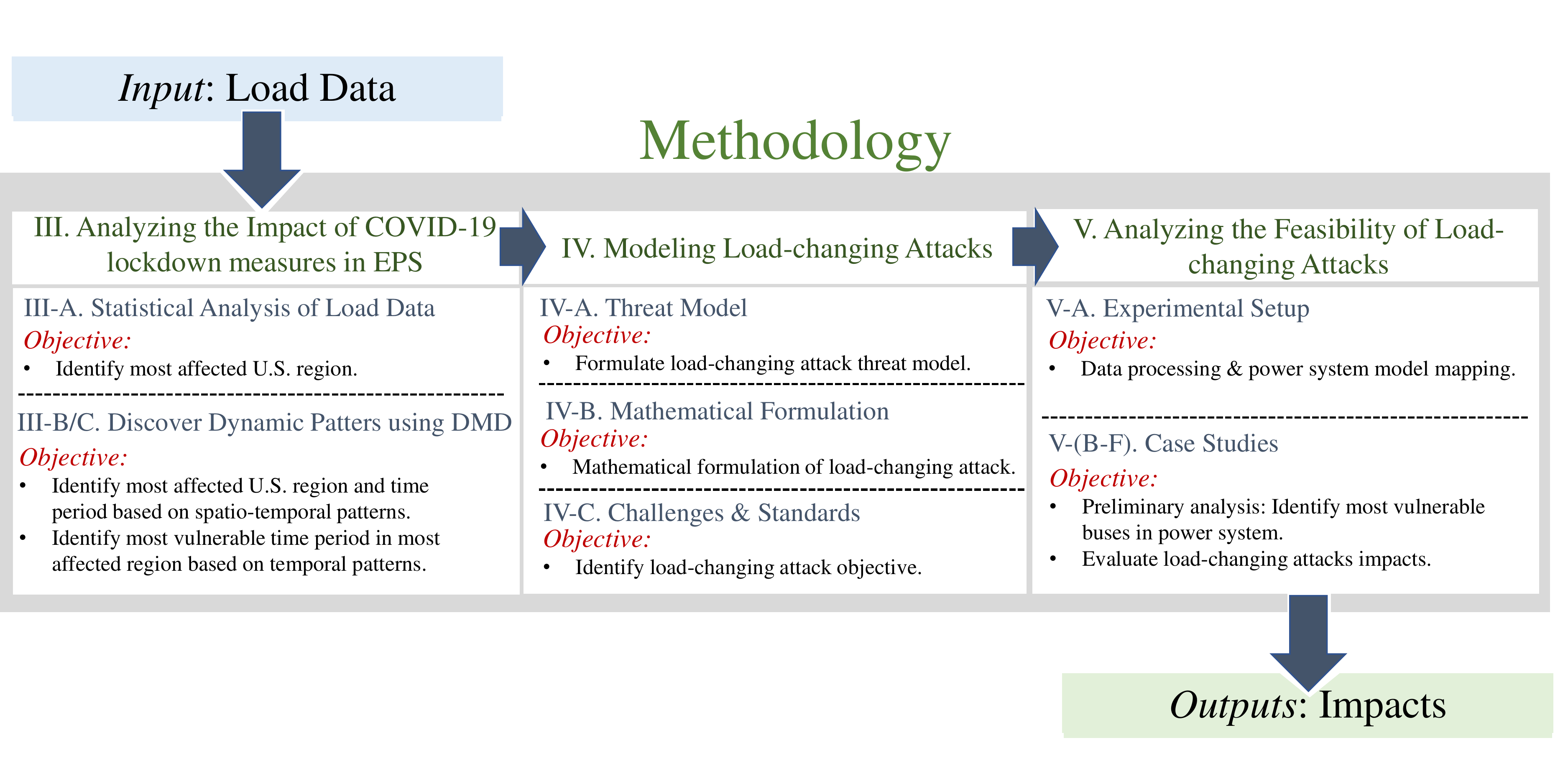}
  \caption{Methodological approach for analyzing the feasibility of load-changing attacks.}
  \label{fig:roadmap}
\end{figure*}

\subsection{Contributions \& Methodology}
This paper focuses on analyzing the feasibility of load-changing attacks in power systems during the 2020 COVID-19 pandemic while using the previous year (2019) as a baseline. The methodological approach to investigate the feasibility of load-changing attacks in CPES is presented in Fig. \ref{fig:roadmap}.
The contributions of this paper, highlighting also the differences between the proposed work and the reviewed literature, can be summarized as follows:

\begin{itemize}
\item A methodology for investigating the feasibility of load-changing attacks in CPES during pandemic-type events is proposed as depicted in Fig. \ref{fig:roadmap}. The proposed methodological approach consists of three main steps: \textit{(i)} An analysis of the impact of  the COVID-19 lockdown measures in load consumption, \textit{(ii)} Modeling of realistic high-impact, low-probability load-changing attacks, and \textit{(iii)} A feasibility analysis that evaluates the impact of the modeled load-changing attacks in a compromised CPES.

\item In order to accomplish \textit{(i)}, dynamic mode decomposition (DMD) and statistical analyses of load consumption data during the COVID-19 outbreak are performed. The objective of these analyses is identifying dynamic load patterns that could cause abnormal loading conditions in CPES, and thus create vulnerable circumstances that attackers could use in their advantage to perform load-changing attacks. Using these analyses, we are able to identify the most affected U.S. region and time periods by recognizing spatio-temporal patterns. 

\item In order to realize \textit{(ii)}, mathematical and threat model formulations are designed in order to  methodically examine a realistic high-impact, low-probability threat targeted at creating frequency instabilities. We investigate the possibility of attackers compromising power system stability taking into consideration abnormal loading conditions caused by lockdown measures as the ones observed during the COVID-19 pandemic.

\item In order to accomplish \textit{(iii)}, the analyses and threat model developed are used in conjunction to experimentally examine and simulate load-changing attack scenarios. The investigation demonstrates that such events can adversely impact the frequency of CPES during the decreased load demand conditions caused by COVID-19 lockdown measures.

\end{itemize}

The rest of the paper is organized as follows. In Section \ref{s:covid19data}, the impact of the lockdown measures is evaluated based on the COVID-19 pandemic response timeline using DMD and statistical analyses. The most affected region in the U.S. is identified and then used jointly with the threat model presented in Section \ref{s:attackformulation}. In Section \ref{s:attackformulation}, the threat model of load-changing attacks and the operational standards for frequency stability are presented. Section \ref{s:experimentalsetup} presents the experimental setup and results of the case studies used to evaluate the feasibility of load-changing attacks during low net-load demand periods, as the ones observed during the COVID-19 pandemic. Finally, Section \ref{s:conclusion} presents conclusions and future work.

\section{Analyzing the Impact of COVID-19 Lockdown Measures in EPS throughout the U.S.}\label{s:covid19data}

\begin{figure*}[t]
\centering
  \noindent\makebox[\textwidth]{\includegraphics[width=18cm]{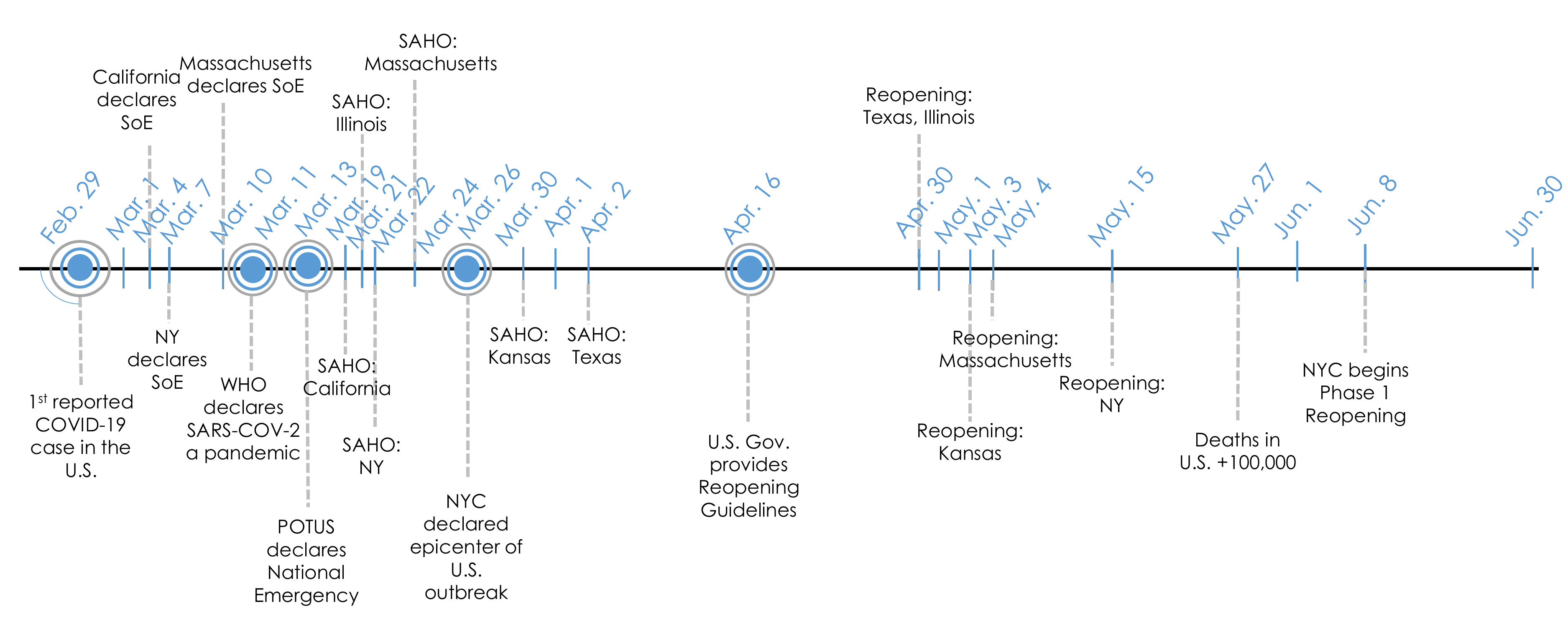}}
  \caption{COVID-19 response timeline from March 1st to June 30th, 2020, with the most important events for the seven states analyzed. SoE - state of emergency, SAHO - stay-at-home-order, WHO - World Health Organization, POTUS - President of the United States.}
  \label{fig:timeline}
\end{figure*}

Before examining the feasibility of load-changing attacks in CPES during the COVID-19 pandemic, it is important to first understand the impact that the lockdown measures, due to the novel COVID-19 pandemic, had in the electricity consumption around the U.S. In order to understand this impact, we obtained and analyzed electricity consumption data obtained from a cross-domain open-source data hub, COVID-EMDA \cite{ruan2020cross}. This data hub integrates weather, satellite imaging, mobile device location, and electricity market data from seven regions that cover some of the top hardest-hit states in the U.S, i.e., California, Texas, 
Midcontinent Central region, Kansas, Illinois, New York, and Massachusetts. In this paper, we focus our attention in analyzing seven main representative urban areas that belong to the major RTOs in the U.S. 
Urban areas have been the most affected areas by COVID-19 spread and lockdown measures; according to the United Nations (UN), they have become the epicentre of the pandemic with more than 90\% of all reported cases \cite{UnitedNations, paul2020progression}. The representative urban areas analyzed and their corresponding RTO are:

\begin{enumerate}
    \item Houston  - ERCOT
    \item Boston - ISO-NE
    \item Central - MISO
    \item New York City (NYC) - NYISO
    \item Chicago - PJM
    \item Kansas City - SPP 
    \item Los Angeles - CAISO 
\end{enumerate}

\subsection{COVID-19 Outbreak Timeline \& Quantification of Load Demand Changes due to COVID-19: 2019 vs. 2020 }

In order to understand the load demand changes and the magnitude of the COVID-19 impacts across the seven major regions examined, we first analyze the possible reasons of why such changes occurred. Thus, we compiled a series of important events, in a timeline, that correlate with the load variation during the outbreak. Fig. \ref{fig:timeline} shows the timeline for the different regions analyzed. As seen in this timeline, the majority of the lockdown measures in the analyzed states occurred during the March-to-May period of 2020. Also, it should be noted that re-openings in most of these places occurred during the second and third weeks of May. Based on this timeline, we selected the period ranging from March 1st to June 30th as the one to compare with the previous year (2019) for load consumption comparisons and impact analysis.

\begin{figure*} \centering    
\subfigure[] { \label{fig:5}     
\includegraphics[width=5.72cm]{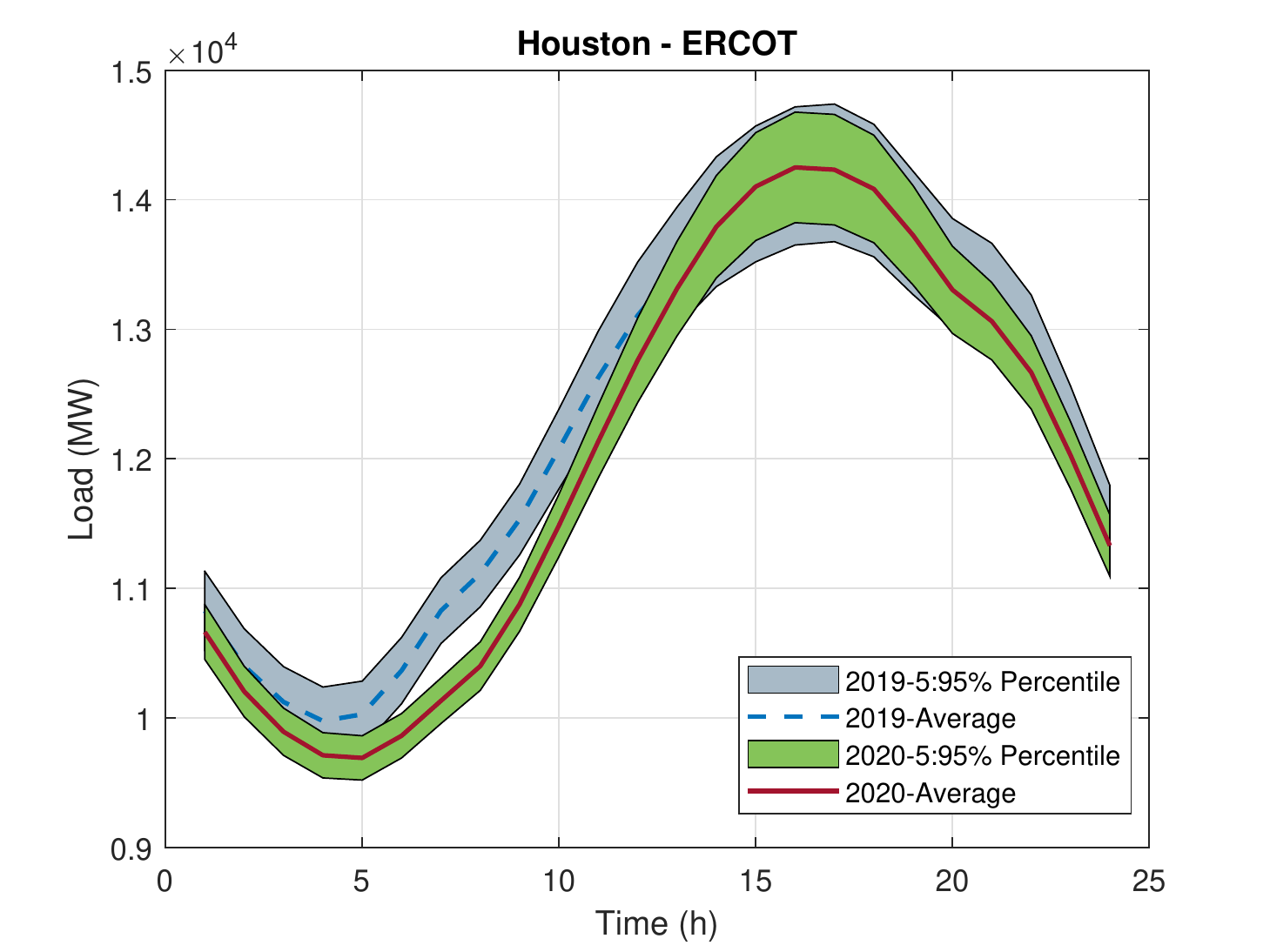}}
\subfigure[] { \label{fig:6}     
\includegraphics[width=5.7cm]{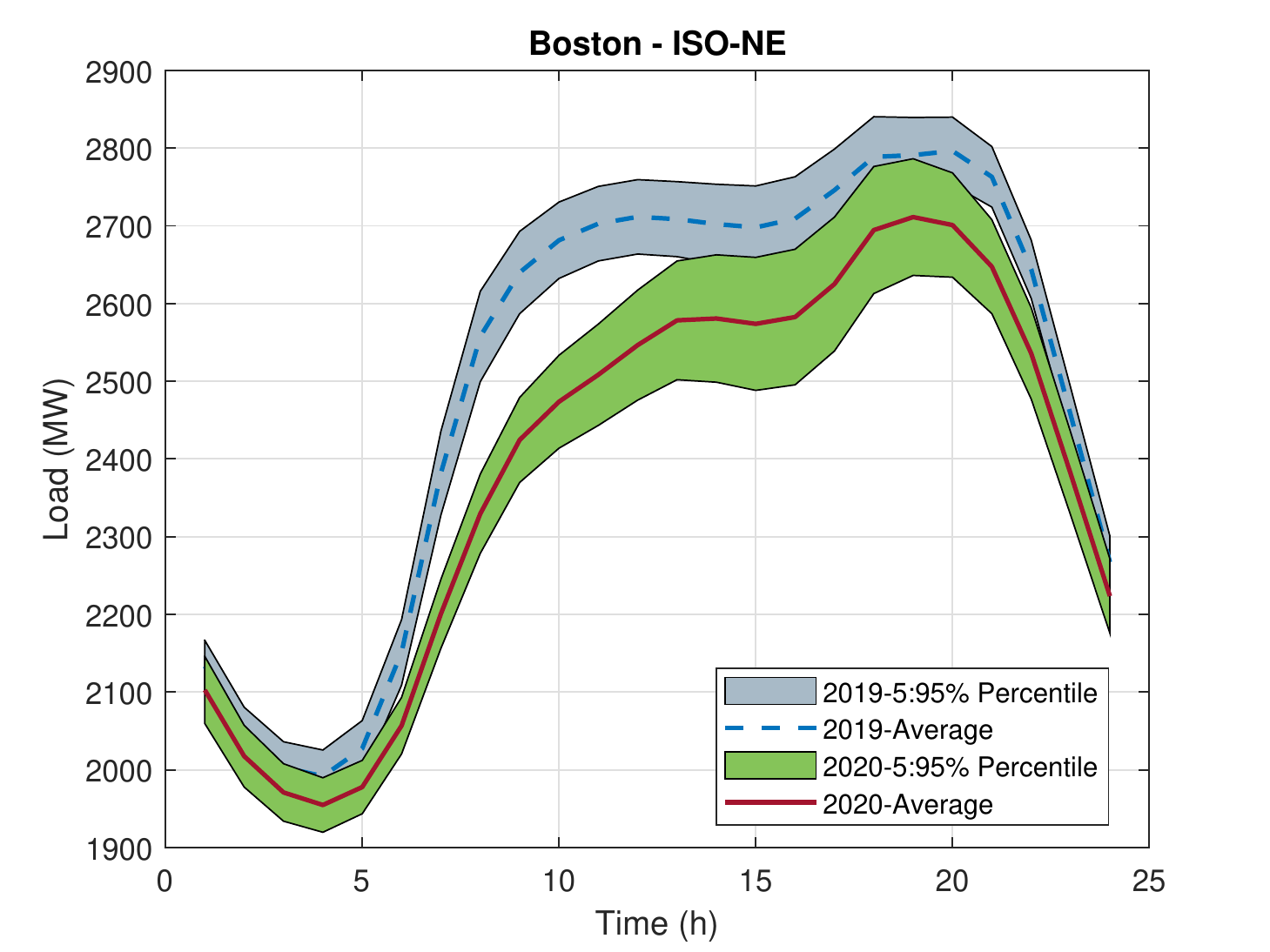}}
\subfigure[] { \label{fig:7}     
\includegraphics[width=5.7cm]{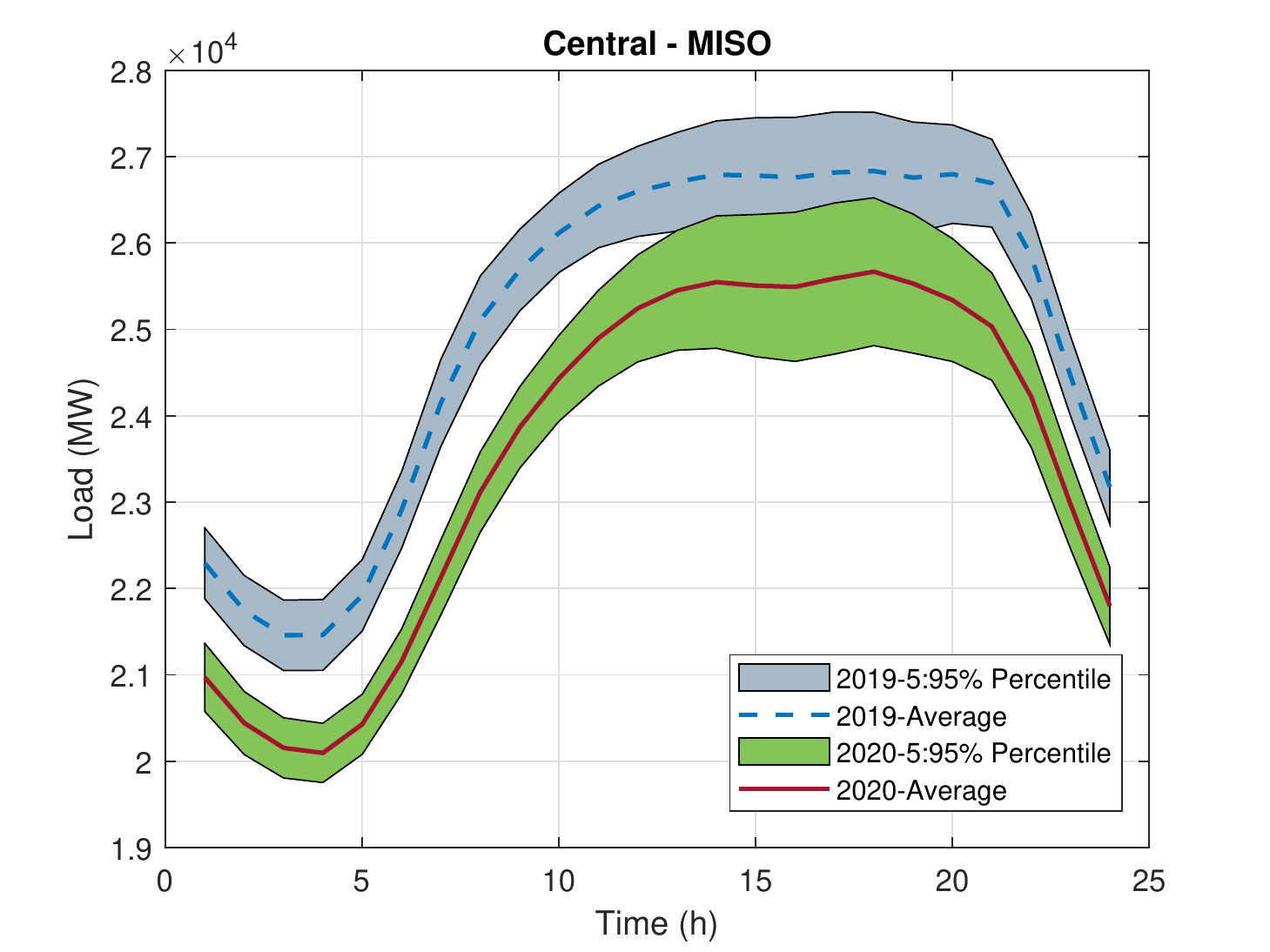}}
\\
\subfigure[] { \label{fig:8}     
\includegraphics[width=5.7cm]{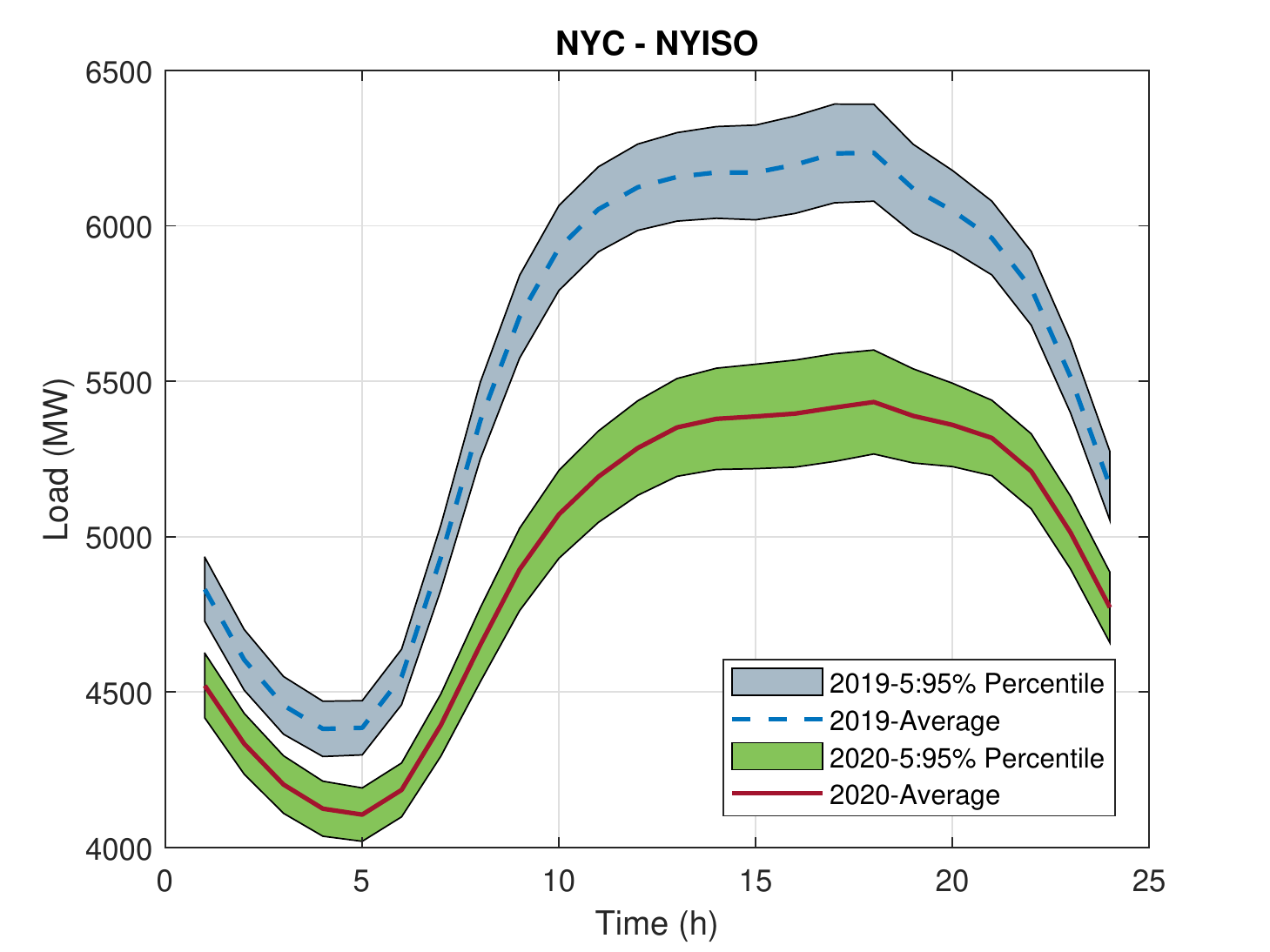}}
\subfigure[] { \label{fig:9}     
\includegraphics[width=5.7cm]{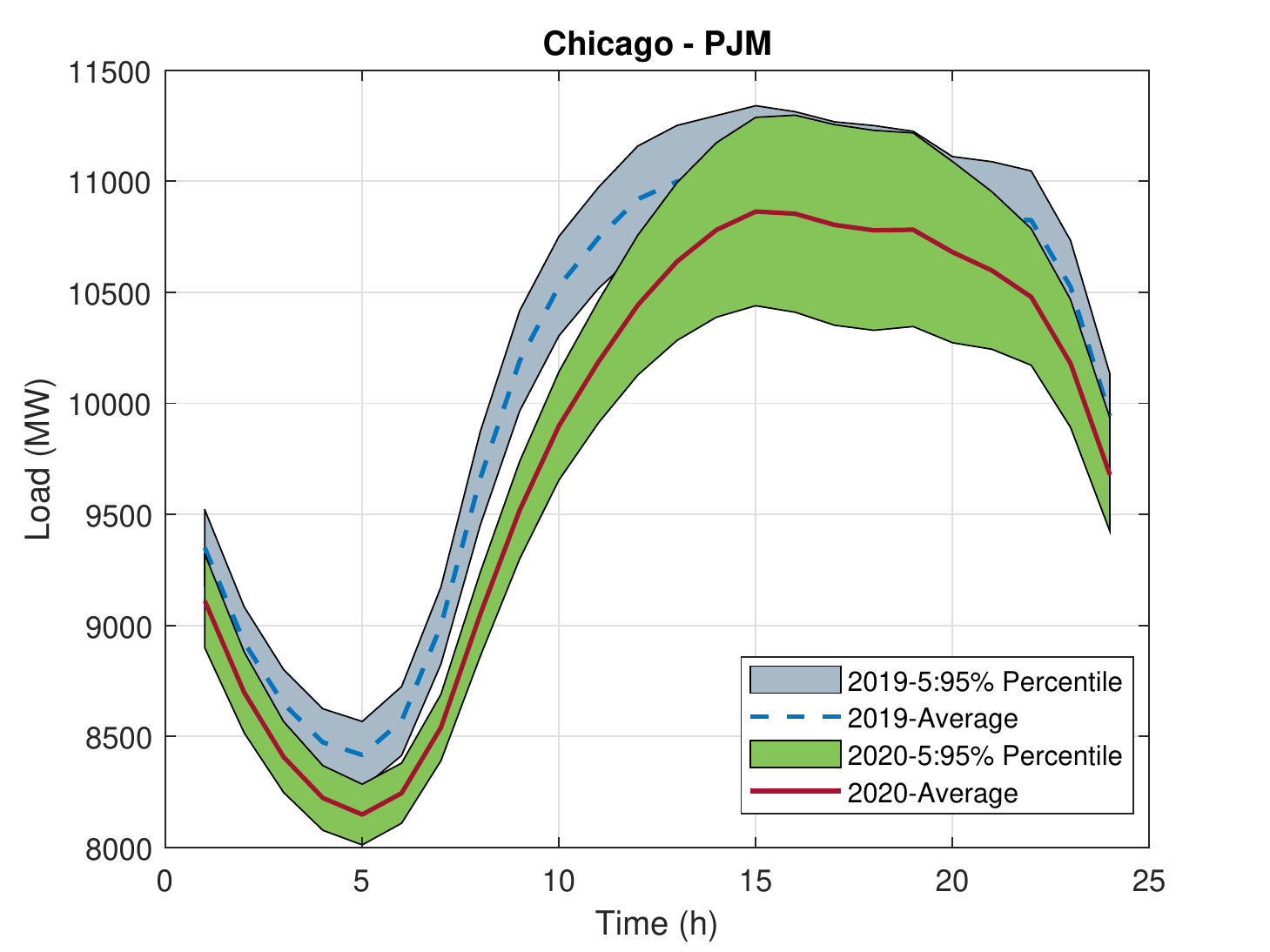}}
\subfigure[] { \label{fig:10}     
\includegraphics[width=5.7cm]{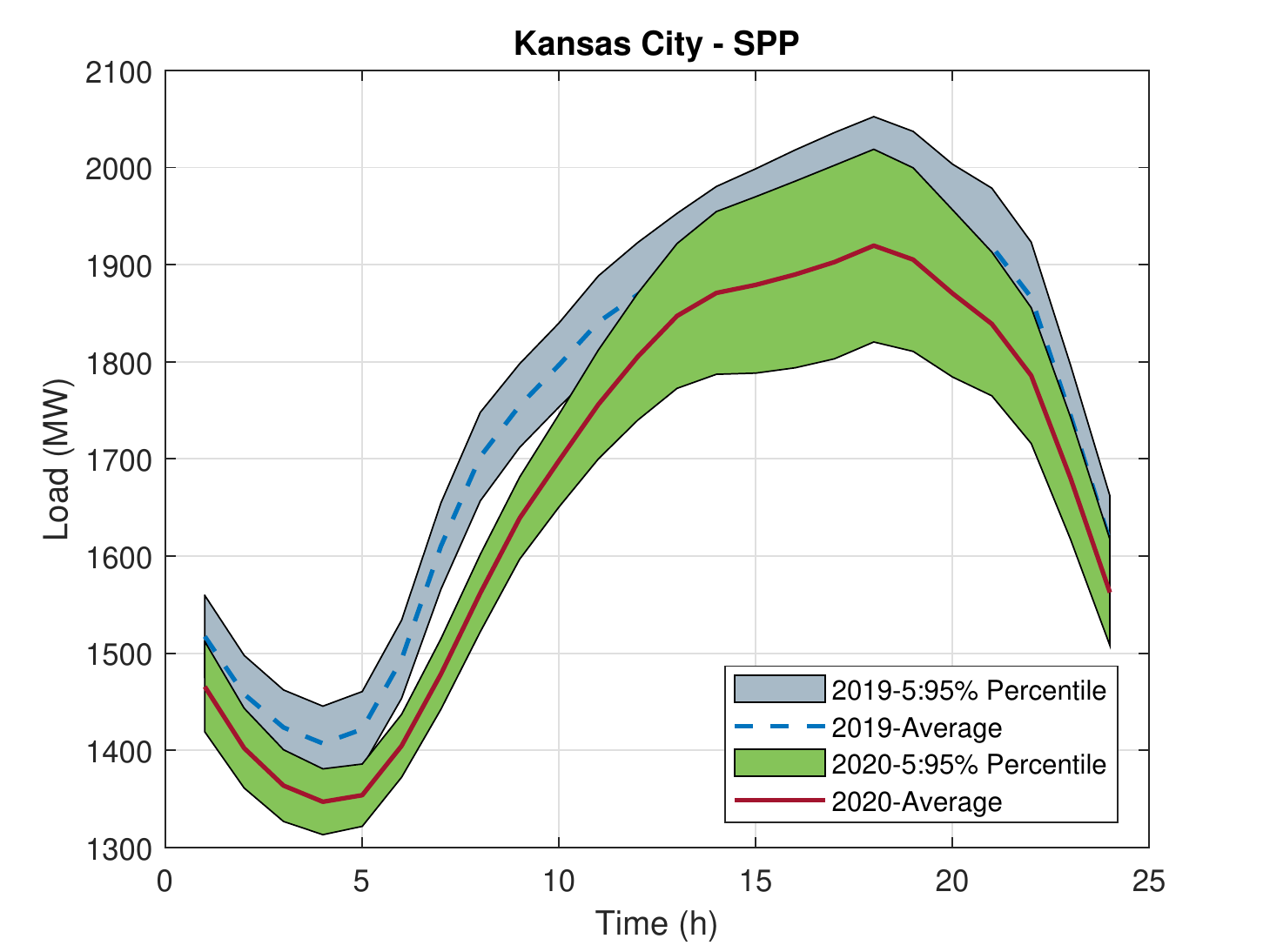} }
\\
\subfigure[] { \label{fig:11}     
\includegraphics[width=5.7cm]{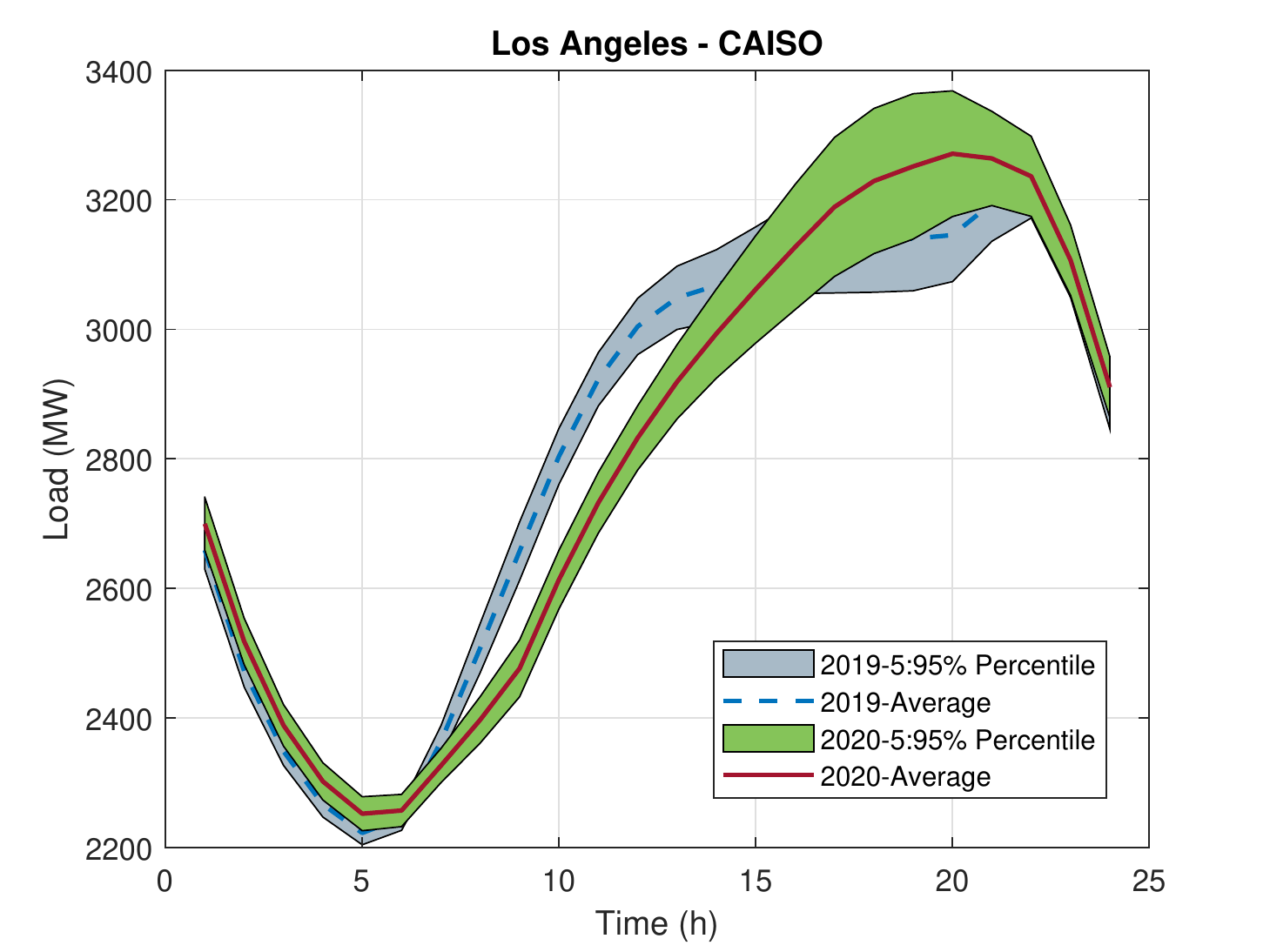} }

\caption{2019 and 2020 load demand 5\% to 95\% percentiles for: a) Houston - ERCOT, b) Boston - ISO-NE, c) Central region - MISO, d) New York City (NYC) - NYISO, e) Chicago - PJM, f) Kansas City - SPP, and g) Los Angeles - CAISO.} 
\label{fig:percentiles}  
\end{figure*}

In Fig. \ref{fig:percentiles}, we can observe how the electricity consumption in the major U.S. zones and RTOs during the COVID-19 pandemic of 2020 was notably reduced, compared to 2019, during the period of March 1st to June 30th. The most affected time periods are the ones  between 6 am and 12 pm in all regions. Also, the most affected city due to the lockdown measures is clearly NYC since a significant load demand reduction can be observed at all times during the lockdown period. In regions such as Central-MISO and NYC-NYISO, overall low net-load demand conditions can be observed throughout the day. In NYC, the highest load difference between the average load of 2019 and 2020 is found at the time period of 11 am to 12 pm, where a load difference of around 860 MW exists (around 10\% of the total maximum load). Similarly, in Central-MISO, the highest load difference between the average load of 2019 and 2020 is found at the time period of 7 am to 8 am, where the load difference is around 2,020 MW (around 5\% of the total maximum load). These results indicate that NYC is one of the most affected regions by the COVID-19 lockdown measures, and thus, we focus our attention on this region. 

\begin{figure*}[t]
\centering
  \noindent\makebox[\textwidth]{\includegraphics[width=18cm]{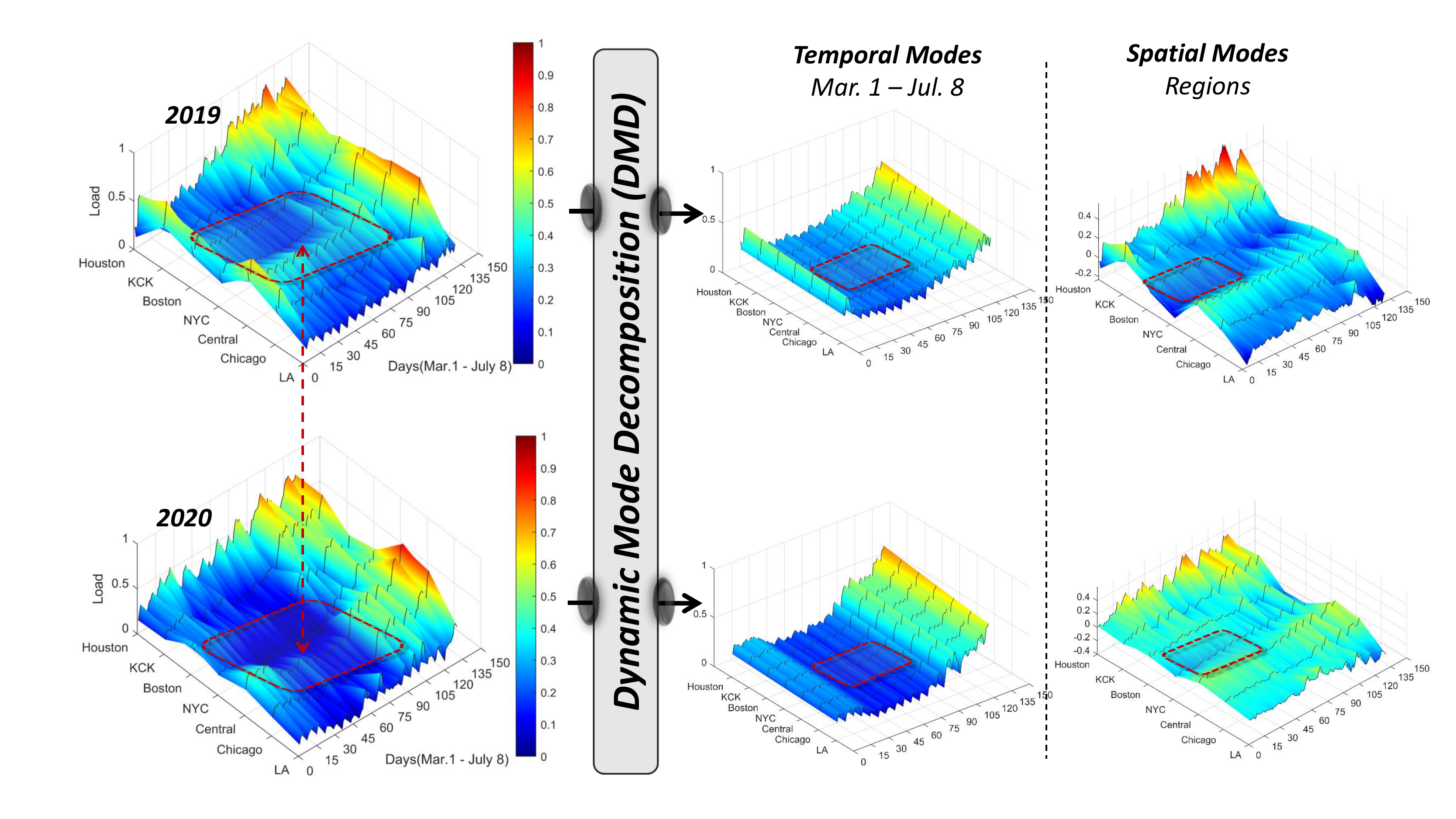}}
  \caption{Surface plots depicting the load consumption data and outputs of the dynamic mode decomposition (DMD) process applied to 2019 and 2020 data during the time-periods of March 1st to July 8th. All surface plots show on the z-axis the load consumption data for all seven regions (CAISO-Los Angeles, PJM-Chicago, MISO-Central, NYISO-NYC, ISONE-Boston, SPP-KCK, and ERCOT-Houston) normalized individually. The x-axis represents the temporal dimension in terms of days during the time period of March 1st to July 8th (130 days) and the y-axis represents the spatial dimension that covers all seven regions, from the first city (Los Angeles) implementing SAHO to the last city (Houston) implementing SAHO according to the COVID-19 response timeline.}  \label{fig:dmdallregions}
\end{figure*}

\begin{figure*}[t]
\centering
  \noindent\makebox[\textwidth]{\includegraphics[width=18cm]{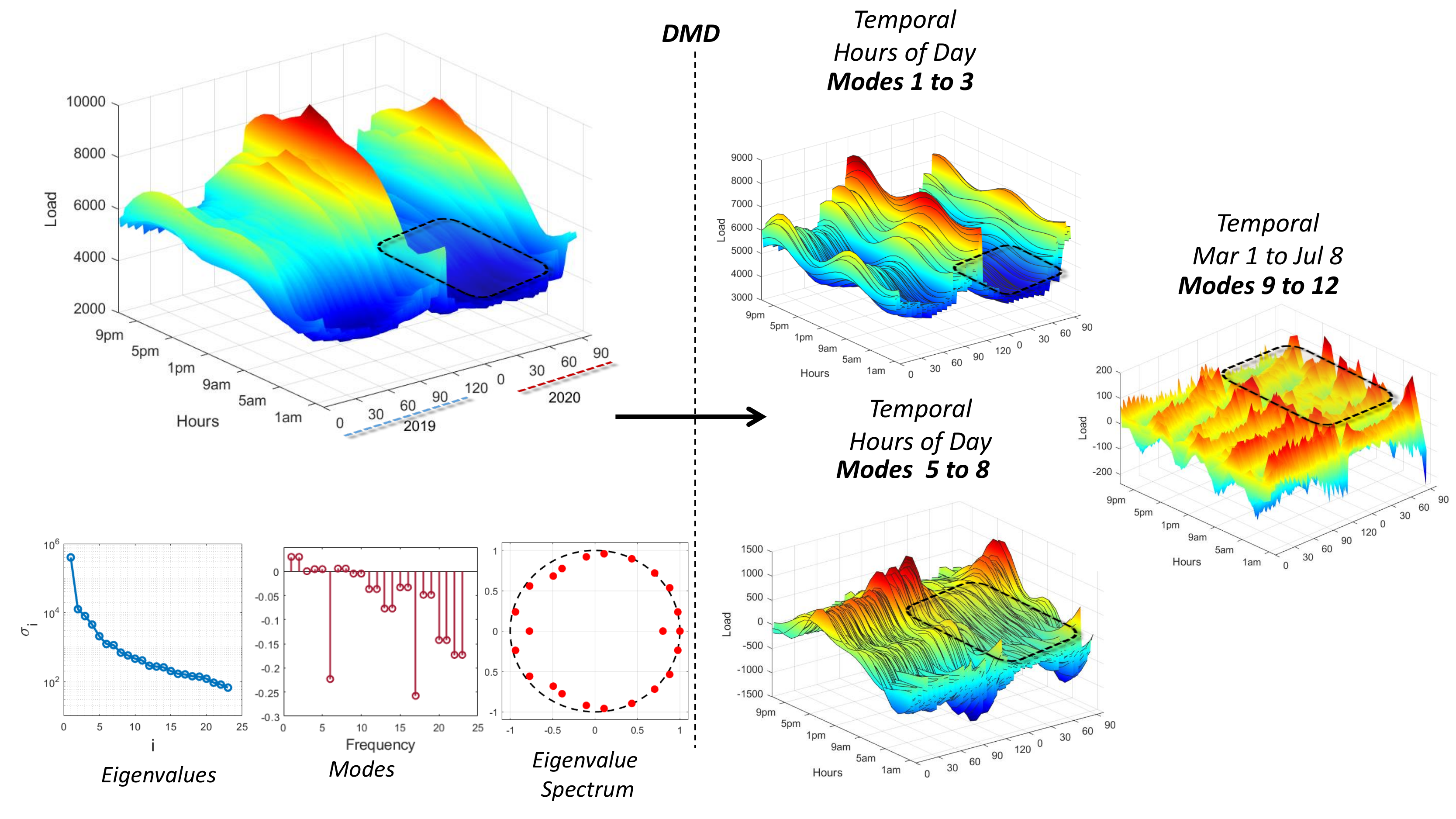}}
  \caption{Surface plots depicting the load consumption data and outputs of the dynamic mode decomposition (DMD) process applied to 2019 and 2020 NYC data during the time-periods of March 1st to July 8th. All surface plots show on the z-axis the load consumption data for NYC. The x-axis represents the temporal dimension in terms of days during the time-period of March 1st to July 8th (130 days) for both the 2019 and 2020 periods combined and the y-axis represents another temporal dimension that covers the different times during the day (12 am to 11pm). The plots shown in the lower left quadrant show the direct outputs of the DMD process, where the first one shows the eigenvalues of $\mathrm{\tilde{A}}$, the second one depicts the modes identified in terms of frequency, and the third one shows the eigenvalue spectrum of $\mathrm{\tilde{A}}$.}
  \label{fig:dmdnyc}
\end{figure*}

\subsection{Discovering Dynamic Patterns in the Reduction of Load Consumption Caused by COVID-19 Across the U.S.}

DMD is a recently developed method capable of performing spatio-temporal decomposition of high-dimensional data \cite{kutz2016dynamic}. This process captures snapshots or measurements in time from a given system and decomposes them into dynamic modes or patterns that can be used to explain the system's behavior. In our case, our `system' is determined by the load consumption data during the time period of the COVID-19 outbreak of some of the most affected regions in the U.S. In order to apply DMD, we structure the load consumption data: $\mathrm{x_{m+1}}$ represents future measurements (snapshots) and $\mathrm{x_m}$ represents previous measurements from the vector $\mathrm{x \in \mathbb{R}^n}$; where $\mathrm{n}$ represents the number of spatial points at each snapshot. For data pairs in $\mathrm{x}$, a best-fit linear operator matrix $\mathrm{A}$ is defined as:

\begin{equation}
\label{eqn:bestfit}
    \mathrm{x_{m+1} = A x_m}
\end{equation}

\noindent where $\mathrm{x_m}$ is equal to $\mathrm{x_m} = [x_1, x_2, ... x_n]$ at snapshot $m$. In our case, the spatial points $n$ represent the individually normalized load consumption values for different affected regions across the U.S. It is important to note that according to \cite{kutz2016dynamic}, the relationship presented in Eq. \eqref{eqn:bestfit} does not need to be exact since other theoretical works have demonstrated the approximation of $\mathrm{A}$ (i.e., the full high-dimensional system matrix) as $\tilde{\mathrm{A}}$ (i.e., the rank-reduced representation) can be used for complex non-linear system applications \cite{proctor2015discovering}. This approximation is useful as it avoids the computational complexity of performing the full eigendecomposition of the high-dimensional system matrix. Using this relationship, we can essentially separate our dynamic system into datasets:

\begin{equation}
\label{eqn:X}
    \mathrm{X = \bigg[ x_1 \ \ x_2 \ ... \ \ x_{m-1} \bigg]}
\end{equation}

\begin{equation}
\label{eqn:X'}
    \mathrm{X' = \bigg[ x_2 \ \ x_3 \ ... \ \ x_m \bigg]}
\end{equation}

\noindent where $\mathrm{X}$ and $\mathrm{X'} \in \mathbb{R}^{\mathrm{n \times m-1}}$. Combining Eq. \eqref{eqn:bestfit}, \eqref{eqn:X}, and \eqref{eqn:X'}, the relationship of the states in our system can be described as:

\begin{equation}
\label{eqn:approxfit}
    \mathrm{X' \approx AX}
\end{equation}

\noindent where the DMD modes, also called dynamic modes of the evaluated dynamical system, are the eigenvectors of $\mathrm{A}$, while each DMD mode corresponds to a particular eigenvalue of the matrix $\mathrm{A}$ \cite{kutz2016dynamic}. However, as mentioned before, in a high-dimensional system, the matrix  $\mathrm{A}$ may be intractable to be analyzed directly. So, in order to avoid the full eigendecomposition of  $\mathrm{A}$, DMD makes use of its rank-reduced representation in terms of the proper orthogonal decomposition-projected matrix, $\tilde{\mathrm{A}}$. All the steps necessary for performing DMD in a given dataset are presented below:

\begin{enumerate}
    \item Perform the singular value decomposition (SVD) of $\mathrm{X}$:
    
    \begin{equation}
        \mathrm{X \approx U\Sigma V^*}
    \end{equation}
    
    \noindent where * denotes the conjugate transpose, and  $\mathrm{U} \in  \mathbb{C}^{n \times r}$, $\mathrm{\Sigma} \in  \mathbb{C}^{r \times r}$, and $\mathrm{V} \in  \mathbb{C}^{m \times r}$. Here, $r$ is the rank of the reduced (truncated) SVD approximation to $\mathrm{X}$. For more details regarding the process and benefits of truncation in DMD see \cite{kutz2016dynamic}.  
    
    \item Calculate the matrix $\mathrm{A}$ using the pseudoinverse of $\mathrm{X}$ obtained using SVD:
    
    \begin{equation}
        \mathrm{A \approx \tilde{\mathrm{A}} = \mathrm{X'V\Sigma^{-1}U^*}}
    \end{equation}
    
    \item Compute the eigendecomposition of $\tilde{\mathrm{A}}$:
    
     \begin{equation}
        \mathrm{\tilde{A}W  = W \Lambda}
    \end{equation}
    
    \noindent where the columns of $\mathrm{W}$ are eigenvectors and $\Lambda$ is the diagonal matrix that contains the respective eigenvalues.
    
    \item Finally, the reconstruction of $\mathrm{A}$ can be performed using $\mathrm{W}$ and ${\Lambda}$, where the eigenvalues of $\mathrm{A}$ are determined using ${\Lambda}$ and the eigenvectors, that represent the patterns of the DMD modes, are determined by the columns of $\Phi$. $\Phi$ is then computed as follows:
    
    \begin{equation}
        \Phi  = \mathrm{X'V\Sigma^{-1}W}
    \end{equation}
    
\end{enumerate}

The utilization of the discussed DMD method can identify coherent spatio-temporal patterns (modes -- $\Phi$) in the dataset by calculating the respective eigenvectors and eigenvalues. Each eigenvalue describes the growth or decay and oscillatory patterns observed in each dynamic mode (eigenvector) identified in the dynamic system. Therefore, DMD is applied to spatio-temporal raw load consumption data, across different regions/cities in the U.S., in order to identify the effects and patterns that the COVID-19 outbreak caused in load consumption reduction across these regions. The $\Delta \mathrm{t}$ chosen are 1-hour and 1-day resolutions so that the variations of the patterns across the selected temporal scales can be easily translated to load consumption variations during the different COVID-19 related events.

The first step performed in our examination is a spatio-temporal analysis designed to capture the variations in load consumption patterns between the years 2019 and 2020. As mentioned previously, we focus on the March-to-July time periods of the aforementioned years since these were the periods when COVID-19-related actions, such as SoE and SAHO, were in full execution. To perform the proposed DMD analysis, we processed and organized the raw load consumption data from the seven different regions (cities) across the U.S. This data is used to create $\mathrm{X}$ and $\mathrm{X'}$ matrices, where $m$ (temporal snapshots) represents different days in the time-period evaluated (March 1 to July 8th), and $n$ (spatial) represents the different regions/cities evaluated (LA, Chicago, NYC, MISO-Central, Boston, KCK, and Houston). 

Fig. \ref{fig:dmdallregions} shows the raw data and outputs of the DMD process executed using the 2019 and 2020 load consumption data, respectively. As seen in this figure, there are significant differences in the load consumption from the years 2019 and 2020 for most of the cities evaluated. By analyzing the temporal and spatial modes identified by the DMD process, we can characterize how the COVID-19 countermeasures impacted load consumption patterns in the seven studied regions. In essence, we can observe, through load consumption data, how COVID-19 was spreading through the U.S. and along the response timeline. In addition, we can also observe that, from the seven cities analyzed, NYC was one of the most affected regions in terms of significant load variation across the U.S. Cross-referencing these results with the timeline presented in Fig. \ref{fig:timeline}, Fig. \ref{fig:dmdallregions} clearly depicts how this low load-valley period is more prominent during the NYC SoE and SAHO events.

The second analysis using the DMD approach is performed based on temporal load consumption data available from NYC. According to the statistical analysis performed in the previous subsection and the spatio-temporal DMD analysis of all the U.S. regions, the NY region was one of the most affected by the COVID-19 lockdown measures in terms of load consumption variation. In Fig. \ref{fig:dmdnyc}, it can be observed how the DMD process is applied in order to identify temporal modes for load consumption during the daily 24 hours and, at the same time, identify the temporal modes in a slower frequency rate (i.e., 1-day resolution) during the March-to-July time periods of 2019 and 2020. The lower left side of the figure shows the eigenvalues, identified modes, and eigenvalue spectrum outputted by the DMD process while the right side shows the reconstructions of some of the most prominent modes that characterize each temporal domain. Noticeable differences between the 2019 and 2020 load consumption can be seen in the figures shown on the right side. For example, based on modes 1 to 3, we observe how the 2020 data is significantly lower (represented with a darker blue) during the morning hours of the day, while modes 5 to 8 show how the evening load is flattened out as the lockdown measures were implemented during the COVID-19 response timeline. On the other hand, the plot reconstructing modes 9 to 12 shows a clear side-by-side comparison that demonstrates how, during the same time period of 2019 vs. 2020, load demand significantly decreased as the COVID-19 outbreak worsened throughout NYC.

\subsection{Impact of COVID-19 Lockdown Measures in NYC Net-Load Demand}

\begin{figure}[t]
\centerline{\includegraphics[width=0.85\linewidth]{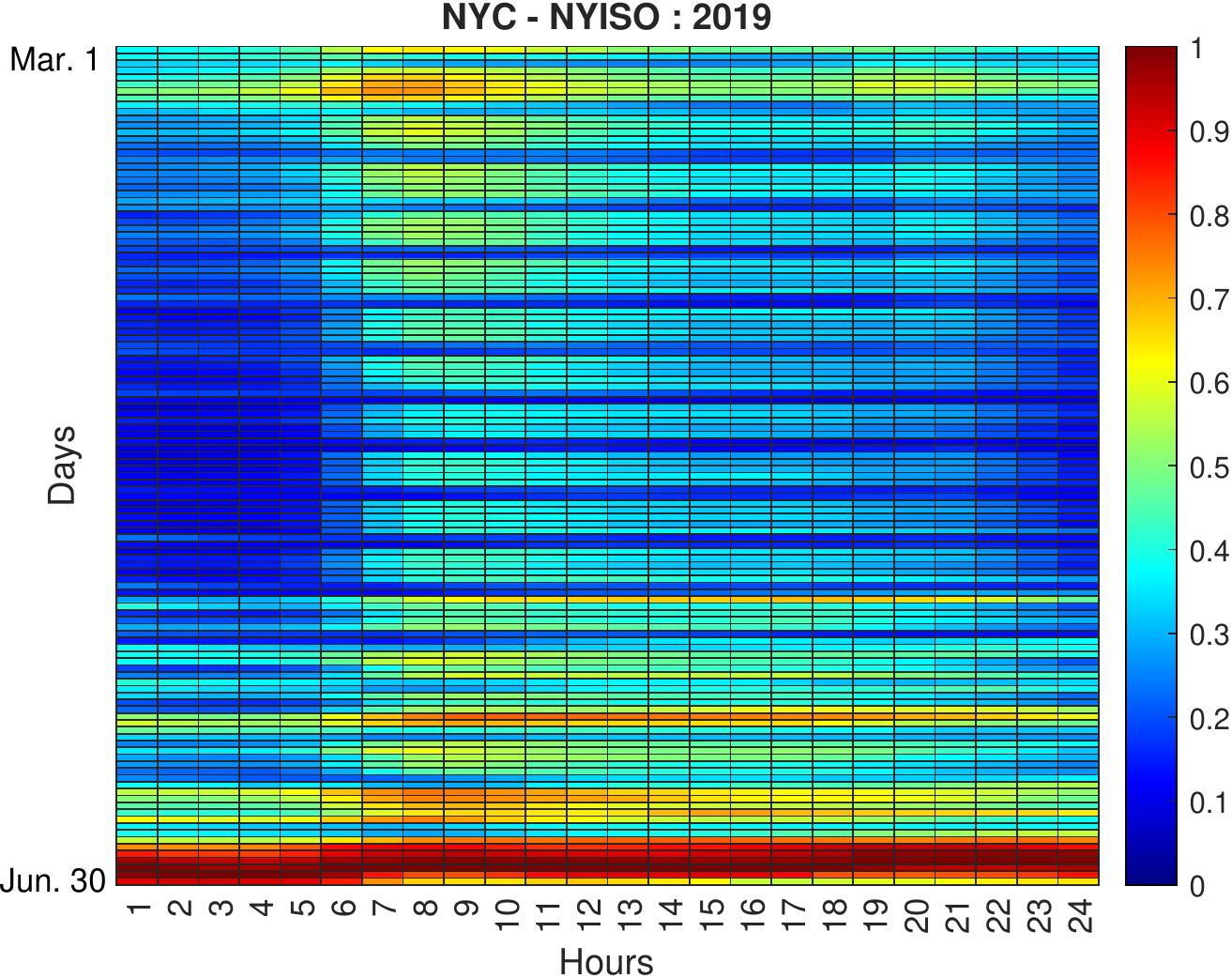}}
\caption{Normalized load demand heatmap for NYC during March 1st to June 30th Period, 2019.} 
\label{fig:heat2019}
\end{figure}

\begin{figure}[t]
\centerline{\includegraphics[width=0.85\linewidth]{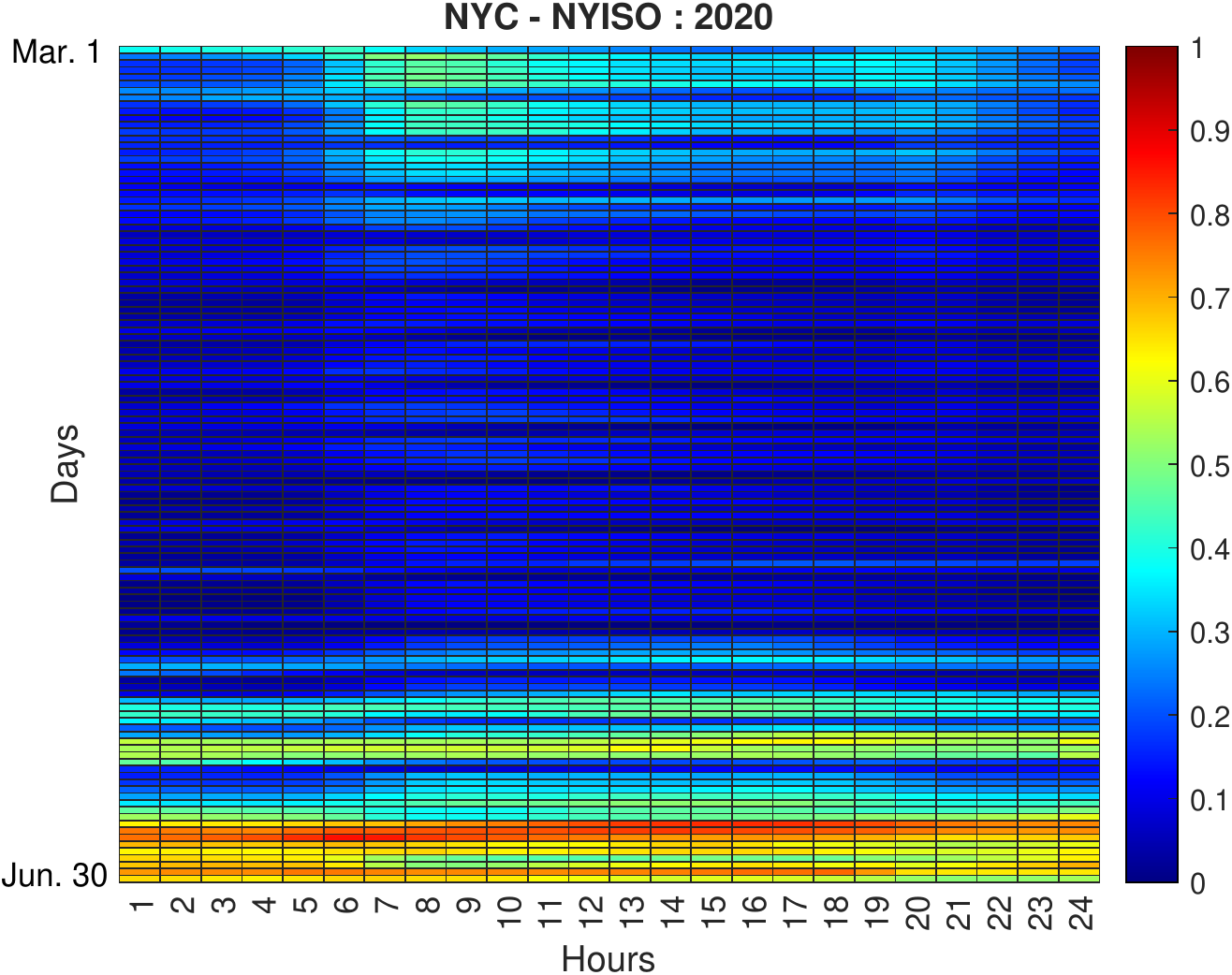}}
\caption{Normalized load demand heatmap for NYC during March 1st to June 30th Period, 2020 (during COVID-19 outbreak).} 
\label{fig:heat2020}
\end{figure}

Based on all the analyses conducted, further investigations are carried out to evaluate the impact of the lockdown measures in NYC, i.e., the most affected city during the analyzed period. Fig. \ref{fig:heat2019} and Fig. \ref{fig:heat2020} show heatmaps comparing the normalized load consumption of NYC during the March 1st to June 30th time period. The vertical represent the different days in the period and the horizontal represent the time of the day. As seen, there is a clear difference between the same periods during 2019 (pre COVID-19 pandemic) and 2020 (during the COVID-19 pandemic). It can also be observed that the most significant variations in load demand reduction are concentrated during the SAHO declared in NYC, that began in March 22 and ended in May 15, as seen in the presented timeline (Fig. \ref{fig:timeline}). The variation observed during this period can be characterized by a 20\% to 30\% reduction in load demand during weekdays.

\begin{figure}[t]
\centerline{\includegraphics[width=0.85\linewidth]{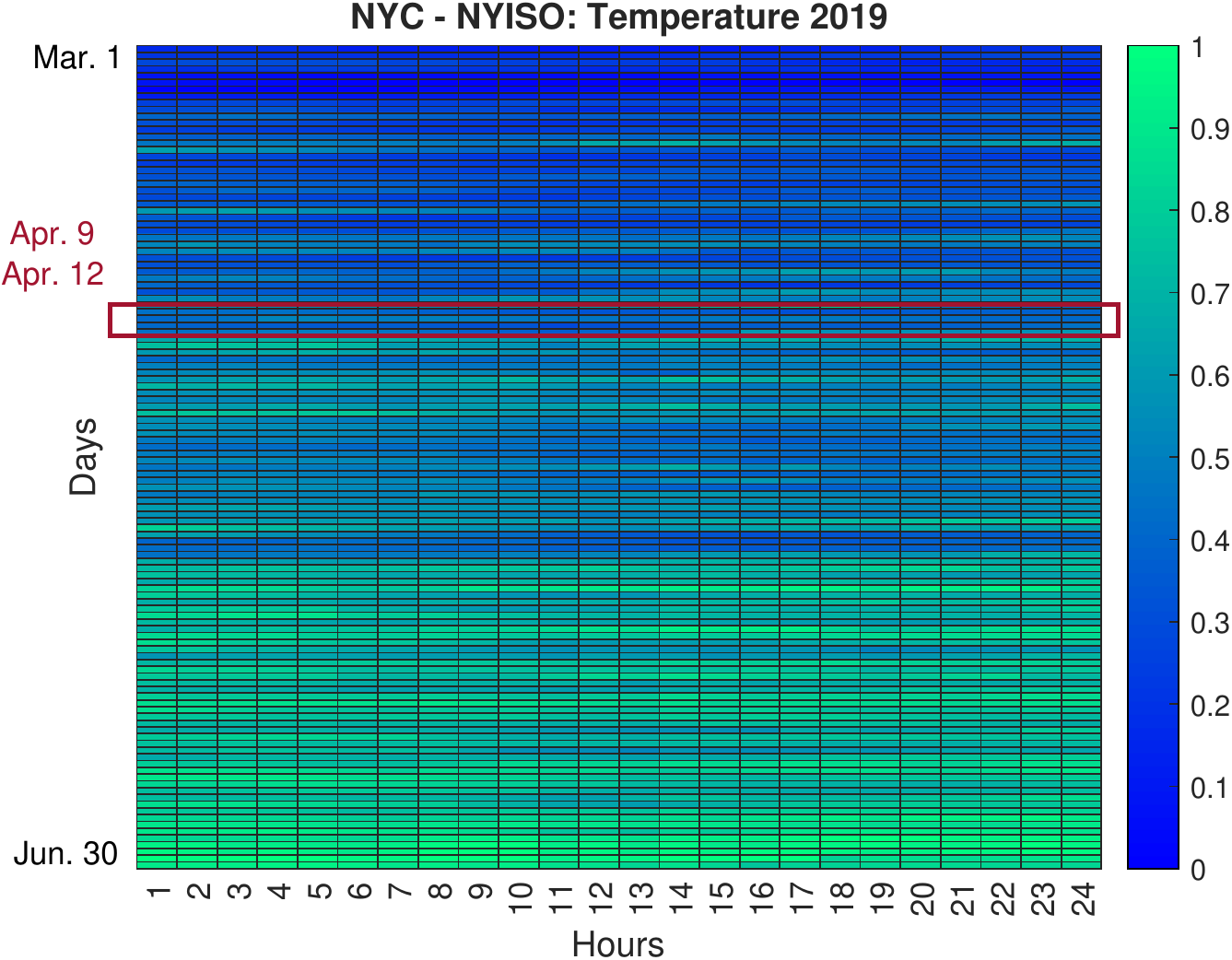}}
\caption{Normalized temperature heatmap for NYC during March 1st to June 30th Period, 2019.} 
\label{fig:heattemp2019}
\end{figure}

\begin{figure}[t]
\centerline{\includegraphics[width=0.85\linewidth]{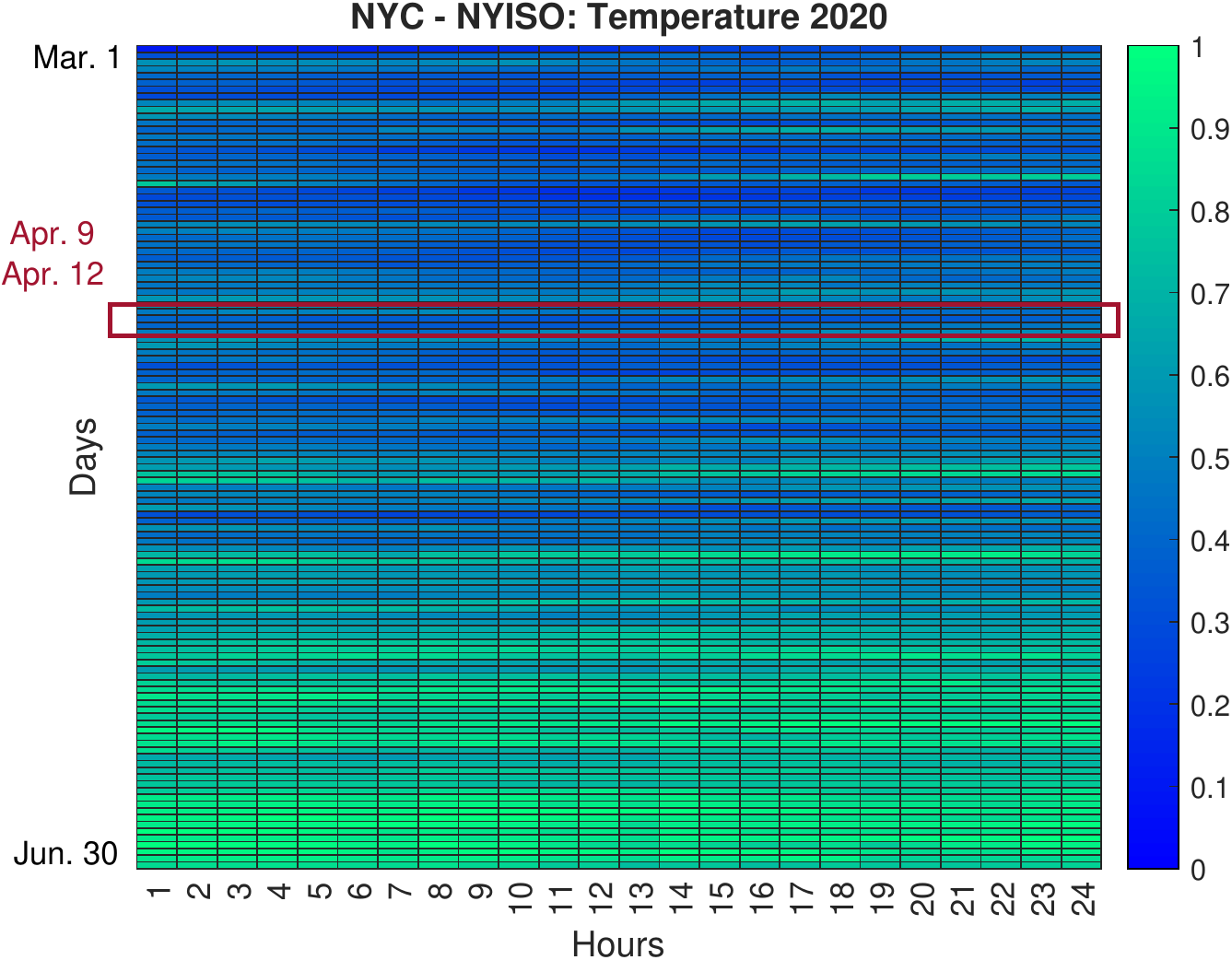}}
\caption{Normalized temperature heatmap for NYC during March 1st to June 30th Period, 2020 (during COVID-19 outbreak).} 
\label{fig:heattemp2020}
\end{figure}

In order to avoid bias from load changes due to non-pandemic related events, we analyzed the weather conditions, and specifically  temperature data for NYC during the same period of time. Figs. \ref{fig:heattemp2019} and \ref{fig:heattemp2020} show the normalized temperature values for NYC during the period of March 1st to June 30th. The temperature data is normalized using -7.22 Celsius as the minimum value and 33.9 Celsius as the maximum value. Based on this data, four days are selected as candidates for investigating the feasibility of load-changing attacks due to low net-load demand conditions, April 9, April 10, April 11, and April 12. These days are selected due to their similarity in temperature values between 2019 and 2020 and falling inside the SAHO time period declared in New York. Choosing these days allowed us to discard temperature as a driving factor of the significant variation in electricity consumption, and helped us to focus on the possible repercussions a pandemic-type event, as the one being experienced, could cause in the cyberphysical security of EPS.

To characterize the temperature similarities between the days during the March 1st to June 30th period, average percentage differences are calculated between the respective days of 2019 and 2020. From the 122 days analyzed in the aforementioned time period, April 11 is under 10\% of the days that had lower average temperature differences between 2019 and 2020. April 12 is under 20\%, April 9 is under 35\%, and April 10 lies under 50\% of the days that had lower average temperature differences between 2019 and 2020. In addition, we also took into account that April 9 and 10 were weekdays while April 11 and 12 from 2020 were weekend days.

\section{Modeling Load-Changing Attacks: Threat Model \& Mathematical Formulation}\label{s:attackformulation}

Modern EPS integrate different technologies, such as intelligent controls and real-time measurement devices, providing system operators with real-time visibility and thus improving system security, stability, and reliability. However, this integration can be a double-edged sword since the use of more interconnected IoT devices exposes the power grid to new cyberattack vectors altering completely the threat model. As more high-wattage loads are equipped with IoT devices, the feasibility of a load-changing attack significantly increases. In this section, we focus on describing load-changing attacks that can be performed by attackers with sufficient motivation, capabilities, and resources to cause major disturbances in transmission and distribution systems via the use of botnets capable of simultaneously controlling large groups of high-wattage loads and DERs.

\subsection{Threat Model: Load-changing Attack } 

In this work, we consider an attacker capable of leveraging vulnerabilities of IoT devices to compromise and control high-wattage appliances, such as HVACs, water heaters, or EV  chargers. By simultaneously switching on and off or granularly adjusting the power consumption of hundreds or thousands of compromised devices, the attacker may be able to cause severe adverse impacts, such as frequency instabilities and over/under voltage conditions, in the CPES. 

Table \ref{tab:threatmodel} shows the threat model used for the load-changing attack according to related work in the area \cite{CPSFrameworkAccesPaper}. As observed, for the load-changing attack, the attacker can be considered as an \textit{oblivious} (i.e., no detailed knowledge of EPS topology) or \textit{semi-oblivious} (i.e., has limited information of the EPS topology) adversary. In addition, since the attack can be performed through IoT-connected devices, \textit{no possession} is required. This means that the attacker does not needs to physically possess the attacked device(s) since they can be compromised through the communication network. For specificity, the attack is considered a \textit{targeted attack} since the adversary's target are devices (e.g., IoT connected high-wattage loads) capable of directly affecting the power grid and possibly cause instabilities that could lead to blackouts and frequency fluctuations in the system. In layman's terms, specificity relates to how specific the attack is, i.e., \textit{targeted} or \textit{non-targeted}. As for adversary's resources, in the investigated load-changing attack, we consider a \textit{Class II} adversary categorization, where \textit{Class I} represents an adversary that does not need and/or has sufficient resources to carry out very complex attacks without being detected, and \textit{Class II} represents an adversary that needs and/or disposes of sufficient motivation and resources to materialize the coordinated load-changing attack without being easily detected. This \textit{Class II} categorization is assumed due to the complexity related to performing a coordinated load-changing attack capable of simultaneously affecting multiple load zones in the power grid that could cause significant damage in the system. This type of attack differs from the dilettante attack mechanism in \cite{acharya2020public} and requires significant resources and knowledge, such as the appropriate instruments and training for being able to infect and/or compromise multiple high-impact load zones, to be effectively carried out.

In addition,  the load-changing attack is considered to have an \textit{iterative attack} frequency and a \textit{multiple-times} reproducibility in terms of the \textit{attack model formulation} of the threat modeling approach. This means that the attack needs to be performed in an iterative manner, i.e., the adversary must attack multiple loads and iteratively change their set-points in order to accomplish the desired effect of destabilizing the system (\textit{iterative attack}). Also, the attack is considered a \textit{multiple-times} attack since it can be performed or reproduced multiple times before being detected and mitigated by operators. Furthermore, the attack level in the \textit{attack model formulation} is considered as a \textit{Level 1} (L1) or \textit{Level 2} (L2) attack according to the level at which the vulnerable assets (e.g., smart HVACs, IoT-connected motors,  PLCs, HMIs, breakers, controllers, etc.) are compromised. These levels can be at the industrial network layer or the local network layer. The \textit{attack technique} describes the attack method used by the adversary. In the case of the load-changing attack considered, the \textit{attack technique} is assumed to be a modification of control logic or a wired/wireless compromise of the controllable loads that affects the integrity of the data in the system. The load-changing attack is considered as a subset of \textit{data integrity attacks} (DIA) due to the fact that the attack is targeted at affecting the integrity of either the system's measurements (e.g., current, voltage, power, or status measurements) or the system's controls (e.g., power set-points or status control changes, etc.). Finally, the \textit{premise} of the attack is related to the integrity of the cyber-system, i.e., how the attack affects primarily the integrity of the ICT devices that make up the communication network. More details regarding the mathematical formulation of the load-changing attack as a DIA is presented in the next subsection.  

\begin{table}[]
\centering
\caption{Threat model formulation of load-changing attacks.}
\label{tab:threatmodel}
\begin{tabular}{||c|c||}
\hline \hline
\textbf{Threat Model \textbackslash Threat} & \textbf{Load-changing Attack} \\ \hline \hline
\textbf{Knowledge} & \begin{tabular}[c]{@{}c@{}}Oblivious\\ or Semi-Oblivious\end{tabular} \\ \hline
\textbf{Access} & Non-possession \\ \hline
\textbf{Specificity} & Targeted \\ \hline
\textbf{Resources} & Class II \\ \hline
\textbf{Frequency} & Iterative \\ \hline
\textbf{Reproducibility} & Multiple-times \\ \hline
\textbf{Level} & L1 or L2 \\ \hline
\textbf{Asset} & \multicolumn{1}{l||}{\begin{tabular}[c]{@{}l@{}}Smart HVAC, IoT-connected motors, \\  PLCs, EV chargers, water heaters, etc.\end{tabular}} \\ \hline
\textbf{Technique} & \multicolumn{1}{l||}{\begin{tabular}[c]{@{}l@{}}Modify control logic or \\ wireless compromise\end{tabular}} \\ \hline
\textbf{Premise} & Cyber: Integrity \\ \hline \hline
\end{tabular}
\end{table}

\subsection{Formulation of Load-changing Attacks}

The load-changing attack presented in this work can be characterized as a DIA-type of attack. In order to present its mathematical formulation, we first consider a cyberphysical system (CPS) plant described by: 

\begin{equation}
\label{eq:cpsx}
    x(k+1) = Gx(k) + Bu(k)
\end{equation}

\begin{equation}
\label{eq:cpsy}
    y(k) = Cx(k) + e(k)
\end{equation}

\noindent where $x(k) \in \mathbb{R}^{n}$ represents the physical system's states, $u(k) \in \mathbb{R}^{l}$ represents the control variables, and $y(k) \in \mathbb{R}^{m}$ represents the system's measurements. The matrices $G \in \mathbb{R}^{n \times n}$, $B \in \mathbb{R}^{n \times l}$, and $C \in \mathbb{R}^{m \times n}$ represent the system, input, and output matrices, respectively. The system input measurement noise is represented by the term $e \in \mathbb{R}^{m}$. The cyber-system of the CPS can be generally expressed as: 

\begin{equation}
\label{eq:cpscyber}
    u(k+1) = Hy(k)
\end{equation}

\begin{figure}[t]
\centering
\includegraphics[width = 0.49\textwidth]{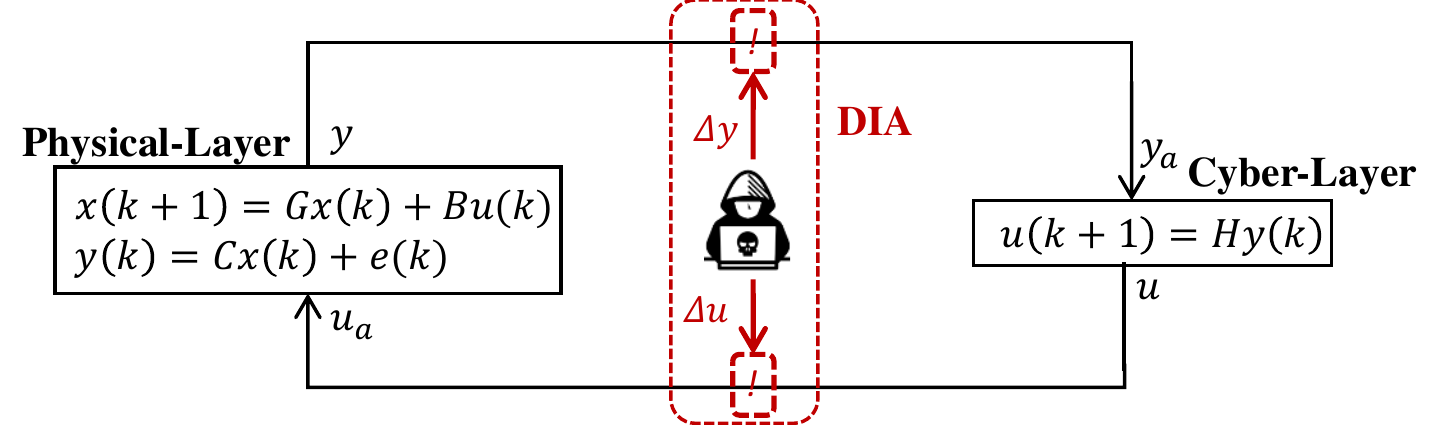}
\caption{\label{fig:cpsplant} Diagram of CPS model and load-changing DIA cyberattack.}
\end{figure}

\noindent $H \in \mathbb{R}^{l \times m}$ represents the control matrix \cite{zhang2018novel}. Fig. \ref{fig:cpsplant} shows a diagram that depicts the variables affected by the DIA load-changing attack in the CPS structure.

As observed in Fig. \ref{fig:cpsplant}, in a DIA, either the measurements ($y$) or the controls ($u$) could be compromised by the adversary via fabrication or modification. More specifically for a load-changing attack, the controls ($u$) of the IoT-controllable loads are `altered/attacked' as:

\begin{equation}
    u_a = u + \Delta u
\end{equation}

\noindent where $u_a$ represents the `altered/attacked' control variables, $u$ represent the original control variables, and $\Delta u$ represent the variations injected by the adversary in the control variables. This modification of controls affects the CPS by:

\begin{equation}
\label{eq:dialoadx}
    x_{a}(k+1) = Gx(k)+Bu_a(k)
\end{equation}

\begin{equation}
\label{eq:dialoady}
    y_{a} = C\big(x(k+1) + B\Delta u(k)\big) + e(k+1)
\end{equation}

\noindent where $y_a$ and $x_a$ represent the input measurements and states of the CPS affected by `altered' control variables.

Mapping the above formulation to a load-changing attack, we can modify the term $u_a$ so it represents `altered' load demand in a CPES as follows:

\begin{equation}
\label{eq:loadchange}
    p_a(k) = p_{i}(k) + \Delta p(k)
\end{equation}

\noindent where $p$ represents the controllable load demand in the system, $p_{i}$ is the initial `un-altered' load demand, $\Delta p$ represents the portion of the total load demand affected by the load-changing attack, and $p_a$ represents the total load demand `altered' by the load-changing attack at one bus. If the attackers simultaneously compromise more than one load/bus in the system, Eq. (\ref{eq:loadchange}) can be extended as:

\begin{equation}
\label{eq:loadchangegeneral}
    P_{T}(k) = \sum_{l=0}^{m} p_{i,l}(k) + \sum_{j=0}^{n} p_{a,n}(k) + P_{loss}
\end{equation}

\noindent where $P_T$ represents the total demand in the system, $m$ is the number of total `unaltered' loads in the system, $n$ is the total number of loads compromised by adversaries, and $P_{loss}$ represents the total losses.

Due to the network power balance, to maintain frequency stability, the sum of all the generation needs to be approximately equal to the total demand and losses in the system: 

\begin{equation}
\label{eq:balence}
    P_{T}(k) \approx \sum_{g=0}^{N_{g}} P_{g}(k)
\end{equation}

\noindent where $N_g$ represents the number of $g$ generators in the system. Hence, in order to understand the effect of load changes in the frequency stability at each generator bus, we can investigate the swing equation  which describes the behavior of rotor dynamics in transient stability studies. The swing equations shown in Eq. (\ref{eq:swing}) - Eq. (\ref{eq:Fequation}) describe the relationship between the input mechanical power ($P_m$), output electrical power ($P_e$), and the rotational speed of the generator ($\omega$) \cite{glover2012power}. The term $P_e$ is directly related to $P_g$ as seen in Eq. (\ref{eq:Pe_V}), since it represents the generator power output plus electrical losses of the generating unit.

\begin{equation}
\frac{2 H}{\omega_{s}} \frac{d^{2} \delta}{d t^{2}}=P_{m}-P_{e}
\label{eq:swing}
\end{equation}

\begin{equation}
\frac{d \delta(t)}{dt}=\omega(t)-\omega_{\mathrm{s}}
\label{eq:FandPA}
\end{equation}

\begin{equation}
\frac{2 H}{\omega_{s}} \frac{d \omega(t)}{dt}=P_{m}-P_{e}
\label{eq:Fequation}
\end{equation}

\begin{equation}
P_e = \frac{V_s V_r}{X} sin(\delta)
\label{eq:Pe_V}
\end{equation}

\noindent $H$ represents the constant normalized inertia, $\omega_{s}$ is the synchronous speed (i.e., 50 or 60 Hz), and $\delta$ is the power angle. $V_s$ is the voltage at the generator bus, $V_r$ is the voltage at receiving bus, and $X$ is the reactance based on the classical model of a generator. The relationship between the electrical frequency $\omega(t)$ with the power angle  $\delta$ is shown in Eq. (\ref{eq:FandPA}). Therefore, it is evident based on these relationships that any sudden change in load demand caused by high-wattage loads turning on/off in the CPES will affect $P_e$, and thus cause frequency fluctuations as seen in Eq. (\ref{eq:Fequation}).

\subsection{Operational Challenges \& Standards: Frequency Thresholds for NERC, ERCOT, and NYISO}

Before analyzing the feasibility of load-changing attacks in a system experiencing low net-load demand conditions such as the ones observed in our investigations, we first explore what are the operational challenges that exist in EPS and are related to load demand changes. According to the Electric Power Research Institute (EPRI) \cite{epricovid}, there are different types of challenging operational conditions, related to net-load demand changes, that could cause steady-state or dynamic threats to the contingency security of EPS. These challenging operational conditions are:

\begin{enumerate}
    \item \textbf{Peak-demand conditions}: Congested networks or limited generation capacity.
    
    \item \textbf{Rapid change in demand or supply conditions}: A rapid change in demand or supply. For example: morning demand ramps, PV-related net demand ramps, or sudden failure of EPS elements. 
    
    \item \textbf{Low net-demand conditions}: Periods when the system load demand is significantly reduced. During these periods, system voltages may rise and system inertia may be affected. Excess generation may be forced to remain online to meet the demand at a later time.
\end{enumerate}

In addition to handling these operational challenges, system operators must provide protective mechanisms to resolve system instabilities, protect system assets, and maintain normal operations. Some of these mechanisms are, for example, underfrequency relays designed to trip when the system's frequency is lower than some predefined values. These predefined values are generally given as frequency bounds programmed into protection mechanisms, and thus are essential to determine the types of remedial actions needed to maintain system stability. The North American Electric Reliability Corporation (NERC) is a nonprofit corporation that provides comprehensive standards for EPS operation in North America. More specifically, NERC requires power systems to operate within a frequency range of 59.5 Hz to 62.2 Hz. If the frequency is out of these bounding ranges, underfrequency or overfrequency protection relays trip parts of the system with the objective of protecting the respective system assets and the overall grid infrastructure \cite{NERC}. Similarly, the Electric Reliability Council of Texas (ERCOT) sets its own frequency thresholds to be 59.3 Hz for underfrequency and 61.8 Hz for overfrequency, respectively. In addition, some system operators may have more complex protection mechanisms that provide rules to shed a certain percentage of the load in the system in case of frequency stability issues \cite{guidessection}.

Since our load data and load-changing attack scenarios are based on NYISO and NYC, we utilize the operational standard (frequency thresholds) provided by  NYISO. In comparison with NERC and ERCOT, NYISO has more strict criteria to mitigate underfrequency and overfrequency scenarios. NYISO defines a major system disturbance as any event that causes the frequency to drop below 59.9 Hz or increase over 60.1 Hz \cite{NYISO}. The specific thresholds given by NERC, ERCOT, and NYISO are provided in Table \ref{table:comparison}.

More specifically, for underfrequency scenarios, NYISO requires fast UFLS to be performed at different percentages when the frequency is rapidly declining. Consecutive 7\% load shedding is performed when the frequency drops below 59.5 Hz, 59.3 Hz, 59.1 Hz, and 58.9 Hz. At this point, if the frequency is still declining, transmission operators must take the necessary steps to minimize damage and service interruption. However, the UFLS required in ERCOT differs from the one applied in NYISO due to its different operational thresholds. In ERCOT, the UFLS starts at 59.3 Hz, where 5\% of the system load is tripped. Then, an additional 10\% of the load is tripped at 58.9 Hz and an additional 10\% at 58.5 Hz. It is important to remember that the intent of UFLS is not to recover the frequency but to stop the frequency decline \cite{ERCOT}.  

On the other hand, for overfrequency scenarios, both ERCOT and NYISO have similar procedures to follow in order to maintain compliance with the NERC Balancing Authority ACE Limit (BAAL) standard. Specifically for the NYISO case, a sustained high frequency of 60.10 Hz is considered an indication of a major load-generation imbalance, and if it continues to decline it can be declared as a `major emergency' \cite{NYISO}. In order to address this emergency NYISO takes the following actions \cite{NYISO}:

\begin{enumerate}
    \item Request all over generating suppliers to adjust their generation and match schedules.
    \item Reduce the applicable dispatchable generation to minimum operating limits.
    \item Request internal generators to voluntarily operate in `manual' mode and below minimum dispatchable levels.
    \item Attempt to schedule variable load or storage to alleviate the problem.
    \item Request reduction or cancellation of all transactions that are contributing to the imbalance event. 
    \item If the overgeneration (i.e., overfrequency) scenario persists, NYISO will declare a `major emergency' and de-commit applicable internal generators until the violation is eliminated. 
\end{enumerate}

\begin{table}[t]
\centering
\caption{Operational frequency thresholds for NERC, ERCOT, and NYISO.}
\label{table:comparison}
\begin{tabular}{||c|c|c|c||}
\hline\hline
\textbf{Category} & \textbf{NERC} & \textbf{ERCOT}& \textbf{NYISO}\\ \hline
Overfrequency (Hz)& 62.20 &61.80& 60.10 \\ \hline
Underfrequency (Hz)& 59.50 & 59.30 &59.90\\ 
 \hline\hline
\end{tabular}
\end{table}

\section{Analyzing the Feasibility of Load-changing Attacks: Experimental Setup \& Results}\label{s:experimentalsetup}

In this section, we present the experimental setup used for evaluating the feasibility of a load-changing attack in a system such as NYISO during low load demand periods such as the ones encountered during the COVID-19 pandemic. As presented in Section \ref{s:covid19data}, four days are selected as candidates for investigating the feasibility of load-changing attacks during low net-load demand periods. These days are April 9, 10, 11, and 12 of 2019 and 2020.

\subsection{Test System: IEEE-14 Bus System}
For our experimental analysis, 5-minute resolution load data are obtained from the 11 load zones that exist in NYISO. Due to the lack of NYISO topological information, we utilize NYISO load data and the respective NYISO load zones are mapped to every load bus in the IEEE-14 bus test system \cite{konstantinou2016case, 8743447}. Below, we describe how to prepare the data in order to examine our load-changing attack case studies. First, the mapping of the NYISO regions to IEEE-14 bus system is performed as follows:

\begin{enumerate}
    \item A - WEST $\rightarrow$ Bus \#2
    \item B - GENESE $\rightarrow$ Bus \#3
    \item C - CENTRL $\rightarrow$ Bus \#4
    \item D - NORTH $\rightarrow$ Bus \#5
    \item E - MHK VL $\rightarrow$ Bus \#6
    \item F - CAPITL $\rightarrow$ Bus \#9
    \item G - HUD VL $\rightarrow$ Bus \#10
    \item H - MILLWD $\rightarrow$ Bus \#11
    \item I - DUNWOD $\rightarrow$ Bus \#12
    \item J - N.Y.C. $\rightarrow$ Bus \#13
    \item K - LONGIL $\rightarrow$ Bus \#14
\end{enumerate}

\begin{figure}[t]
\centering
\includegraphics[width = 0.5\textwidth]{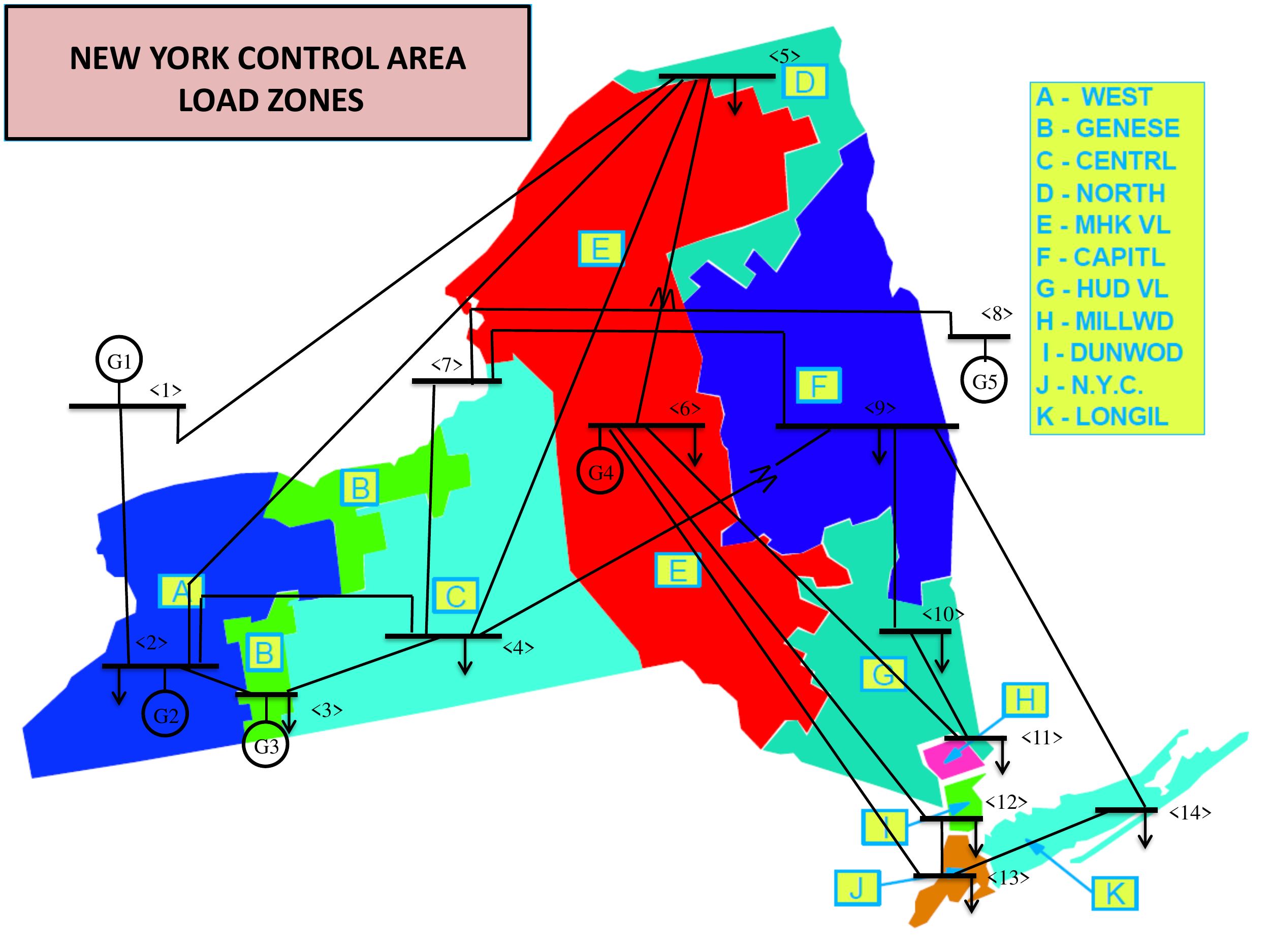}
\caption{\label{fig:nyisomap} NYISO control area load zones mapped to IEEE-14 bus system.}
\end{figure}

\begin{figure}[t]
\centering
\includegraphics[width = 0.48\textwidth]{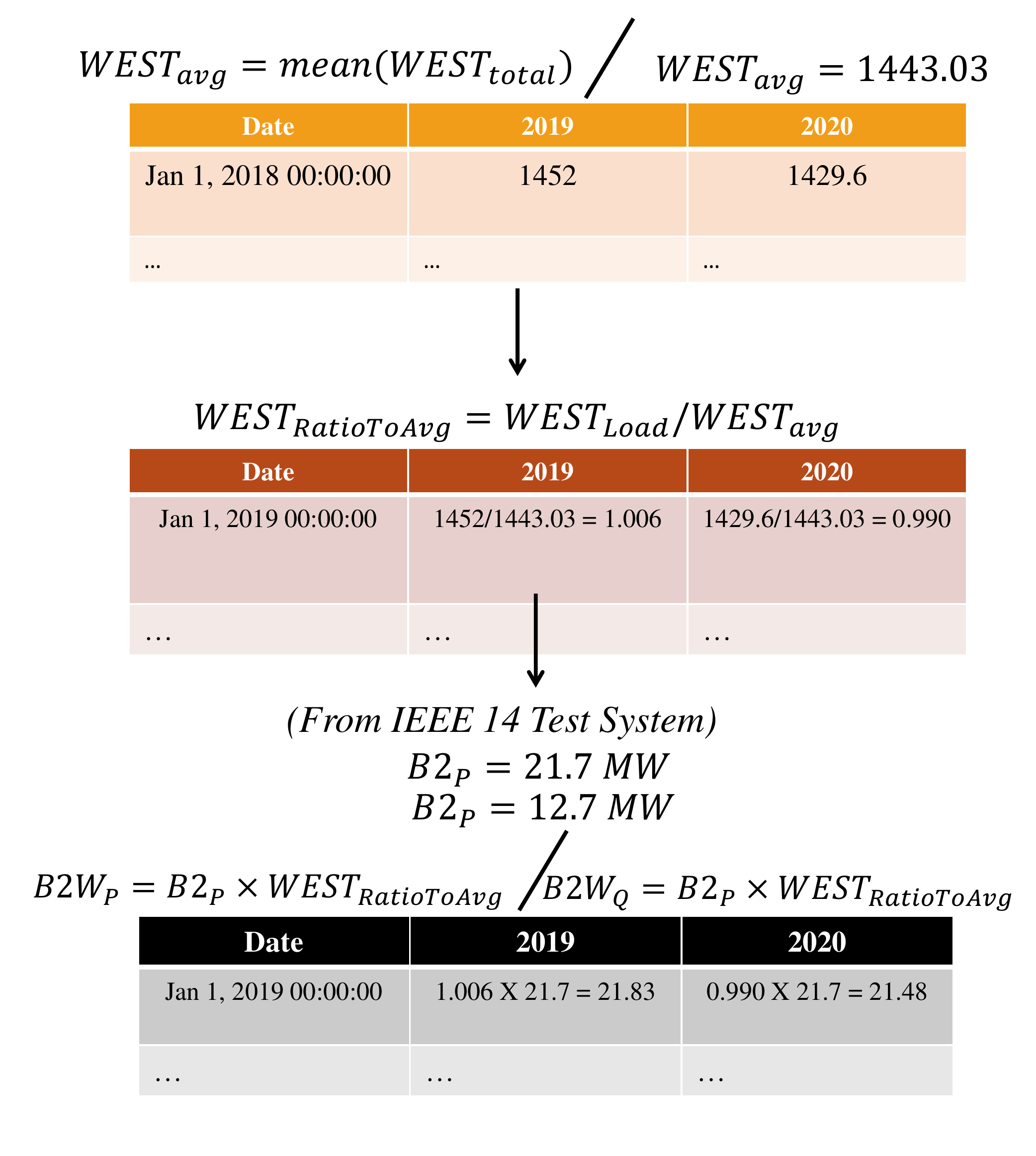}
\caption{\label{fig:process} Normalization process for IEEE 14 test system using NYISO load data.}
\end{figure}

Fig. \ref{fig:nyisomap} shows the NYISO map with the corresponding mappings. In order to adapt the NYISO load values to the IEEE-14 test system, we performed a normalization process that consists of obtaining the average load consumption of each zone (based on historical data), calculating a ratio of load demand, and finally computing the corresponding load value for the IEEE-14 test system. Fig. \ref{fig:process} demonstrates all the steps of the process based on the WEST zone. All other regions follow the same process for computing their corresponding values. 
The adapted version of the IEEE-14 bus test system, with its respective 2019 and 2020 load profiles, is modeled and evaluated using the Power System Analysis Toolbox (PSAT) \cite{PSAT2}. PSAT is an open-source MATLAB toolbox specifically designed to perform power system analysis and simulation.

\begin{figure*} \centering    
\subfigure[] { \label{fig:ldapril9}     
\includegraphics[width=7.72cm]{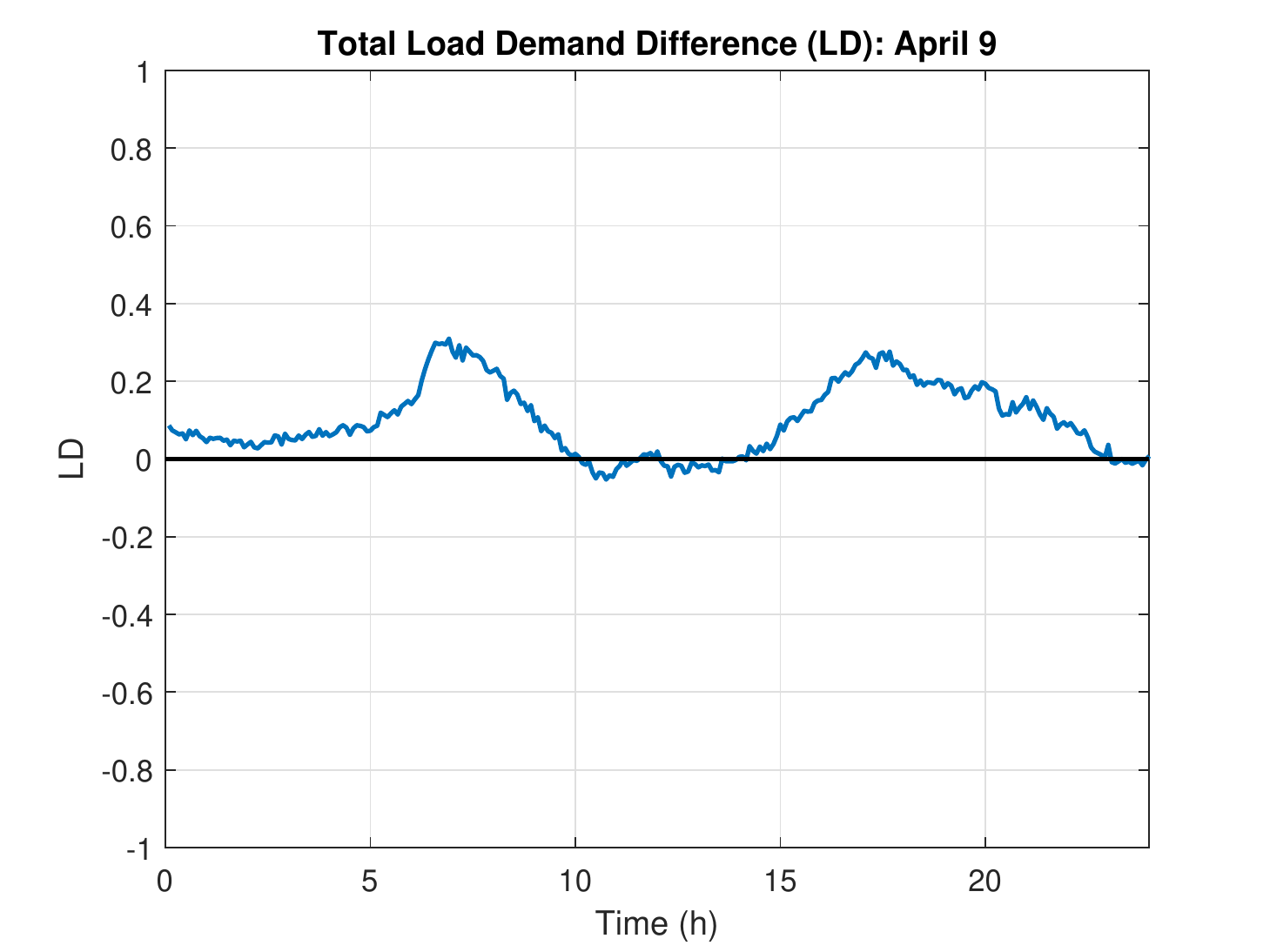}}
\subfigure[] { \label{fig:ldapril10}     
\includegraphics[width=7.7cm]{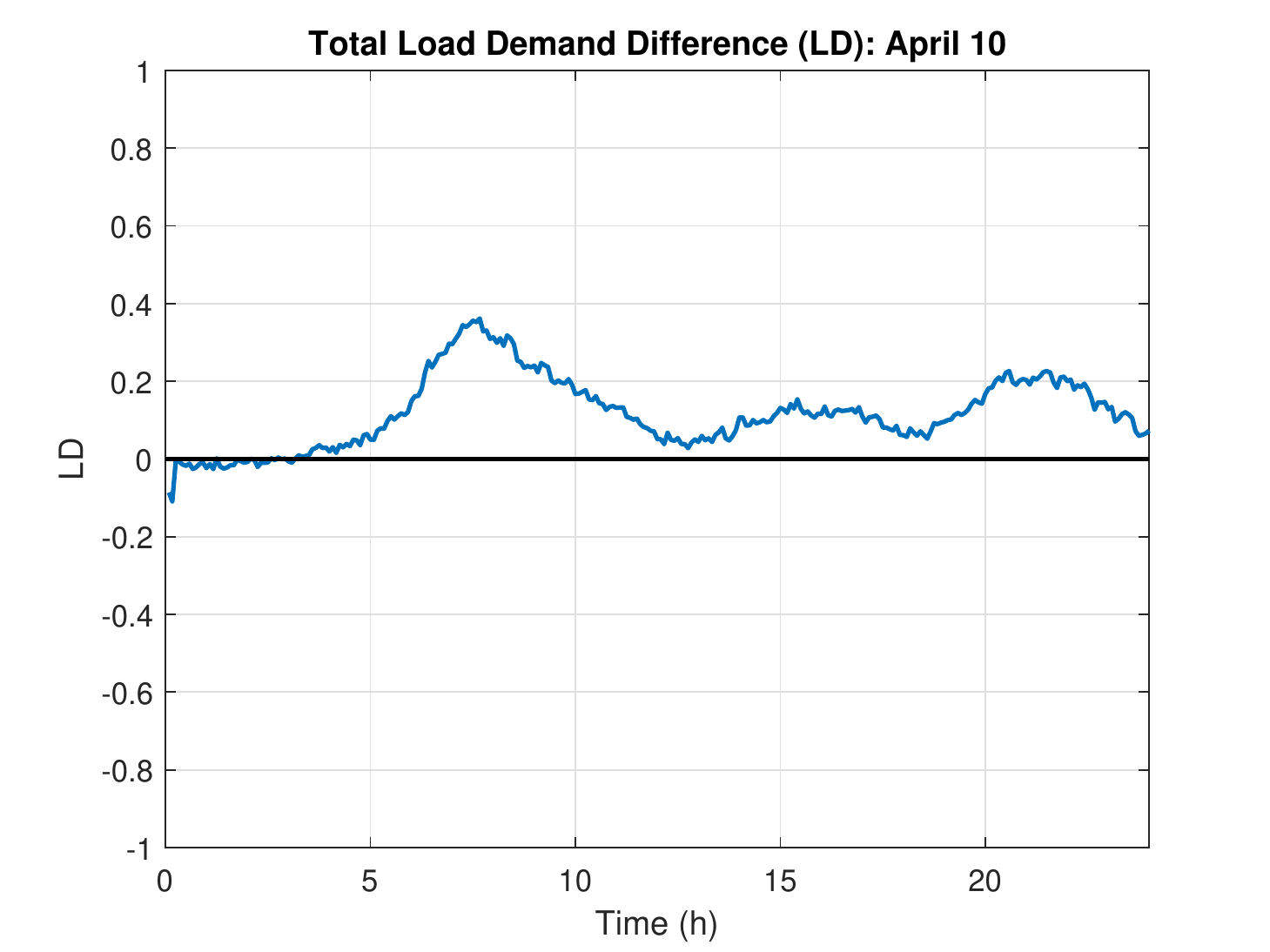}}
\\
\subfigure[] { \label{fig:ldapril11}     
\includegraphics[width=7.7cm]{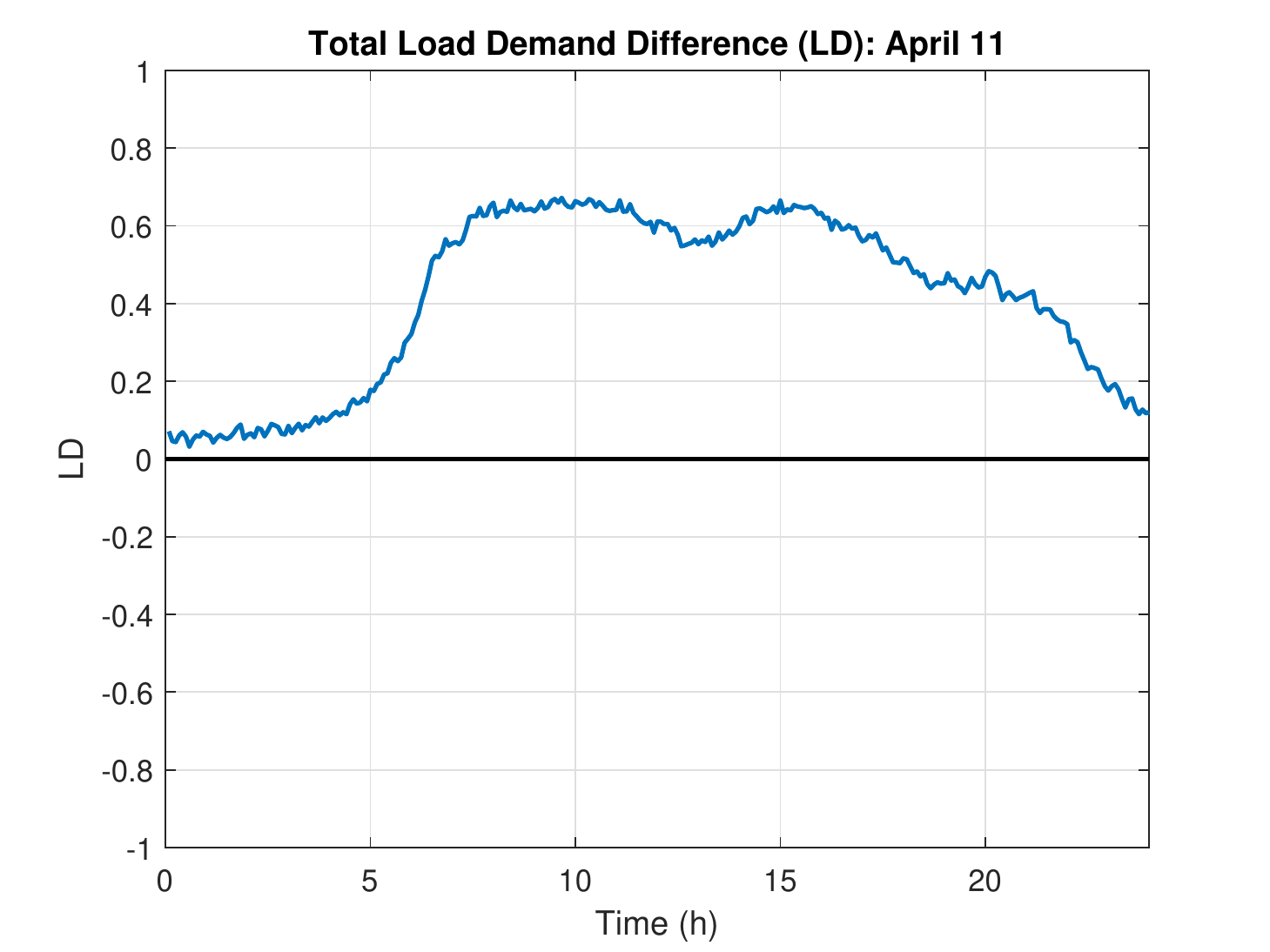}}
\subfigure[] { \label{fig:ldapril12}     
\includegraphics[width=7.7cm]{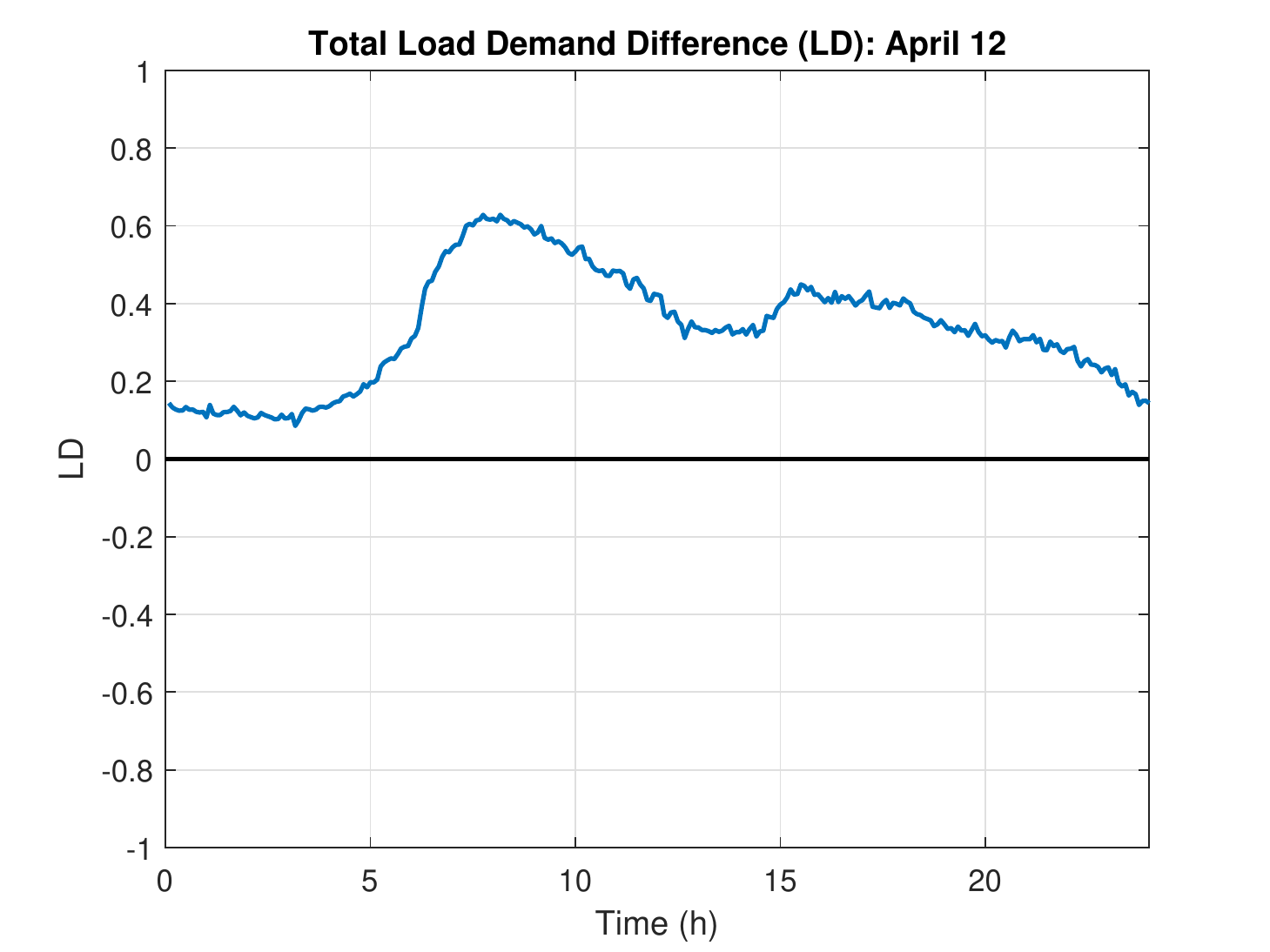}}
\caption{Total load demand difference (LD) calculated for the days: a) April 9, b) April 10, c) April 11, and d) April 12.} 
\label{fig:ldaprildays}  
\end{figure*}

\subsection{Description of Case Studies}

Four 24-hour runs for the days April 9, 10, 11, and 12 are analyzed and inputted into the adapted IEEE-14 bus test system modeled in PSAT. The objective of this study is to find times where the system could be more vulnerable to load-changing attacks due to low net-load demand conditions caused by action events of the COVID-19 pandemic.

For each case study, or day analyzed, a preliminary analysis is conducted to determine the time of day when the 2020 system is more vulnerable (i.e., has lower demand) than the 2019 system, and also to determine the loads (or buses) that have the highest impact in frequency stability of the system. In other words, this analysis indicates the period where the biggest difference in the total load demand of the system exists, when comparing 2019 and 2020 load profiles, and then clearly shows which buses, if compromised, would have the highest impact in the 2020 system. The \textit{total load demand difference} is calculated by subtracting the total load demand of 2019 minus the total load demand of 2020:

\begin{equation}
    LD = TL_{2019} - TL_{2020}
\end{equation}

\noindent where $LD$ is the total load demand difference between 2019 and 2020, and $TL$ is the total load demand of the respective years. 

Additionally, in order to determine the buses (or loads) that, if compromised, would have the highest impact on the 2020 system when compared to the 2019 system, a ratio of the load at each bus and the total load for the specific year (i.e., 2019 or 2020) is calculated. Then, these ratios are subtracted to compute the \textit{load impact index difference} ($LIID$) for each individual bus in the system as seen in Eq. (\ref{loadimpactindex}).

\begin{equation}
\label{loadimpactindex}
LIID_{i} = \frac{L_{2019}^{i}}{TL_{2019}} - \frac{L_{2020}^{i}}{TL_{2020}}  
\end{equation}

\noindent where $L$ is the load at the respective load $i$ for the respective year, 2019 or 2020. It should be noted that all the values used in this analysis are in per unit (p.u.).

After the preliminary analysis is conducted and the `ideal' period of time to attack the system is identified, a load-changing attack is conducted in each selected day and period, while the frequency of the system is monitored. The selection of the compromised bus(es) for each case study depends on the results obtained in the preliminary analysis. More details regarding the specific case studies are presented below.

\begin{figure*} \centering    
\subfigure[] { \label{fig:liidapril9}     
\includegraphics[width=7.72cm]{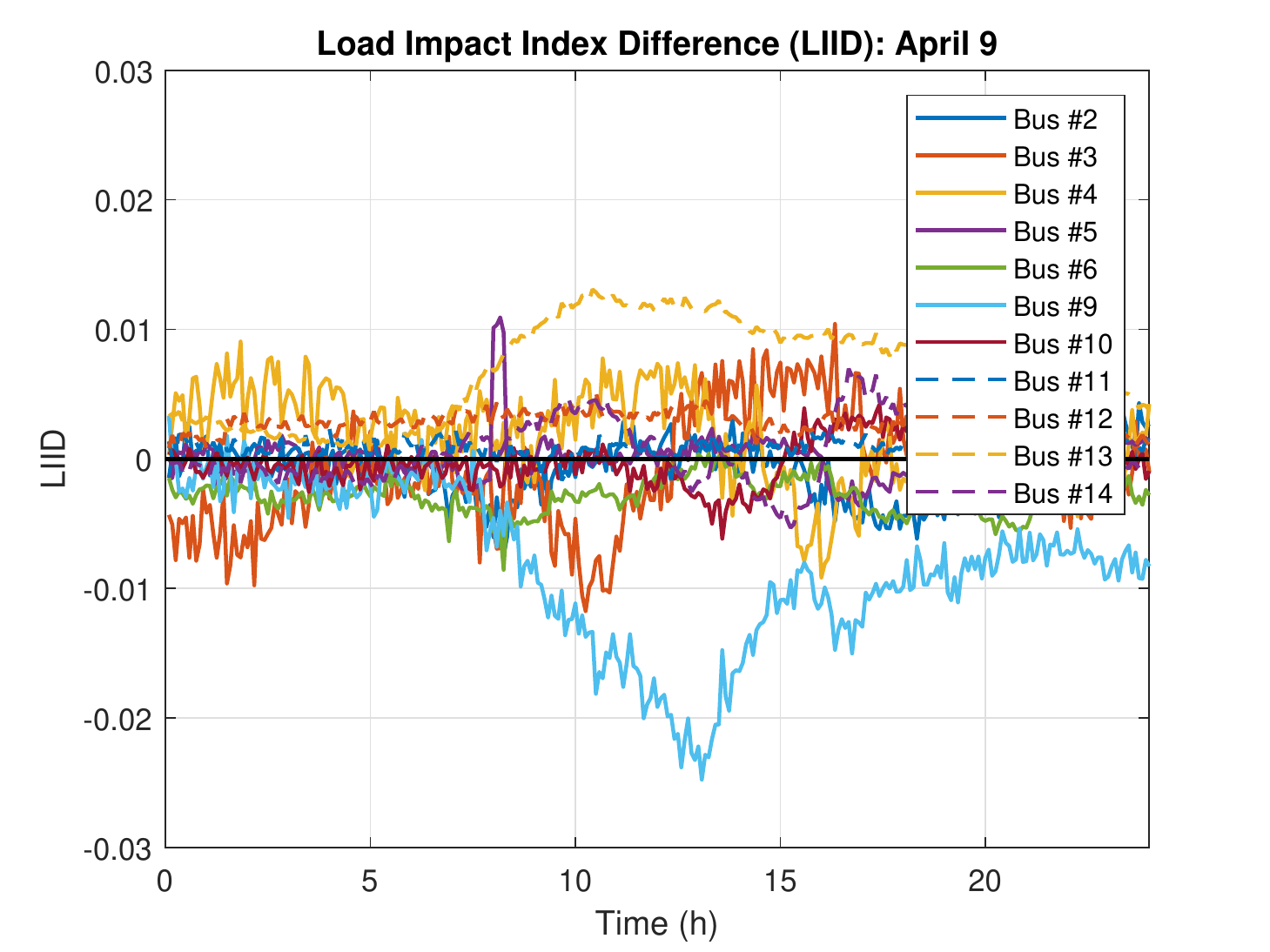}}
\subfigure[] { \label{fig:liidapril10}     
\includegraphics[width=7.7cm]{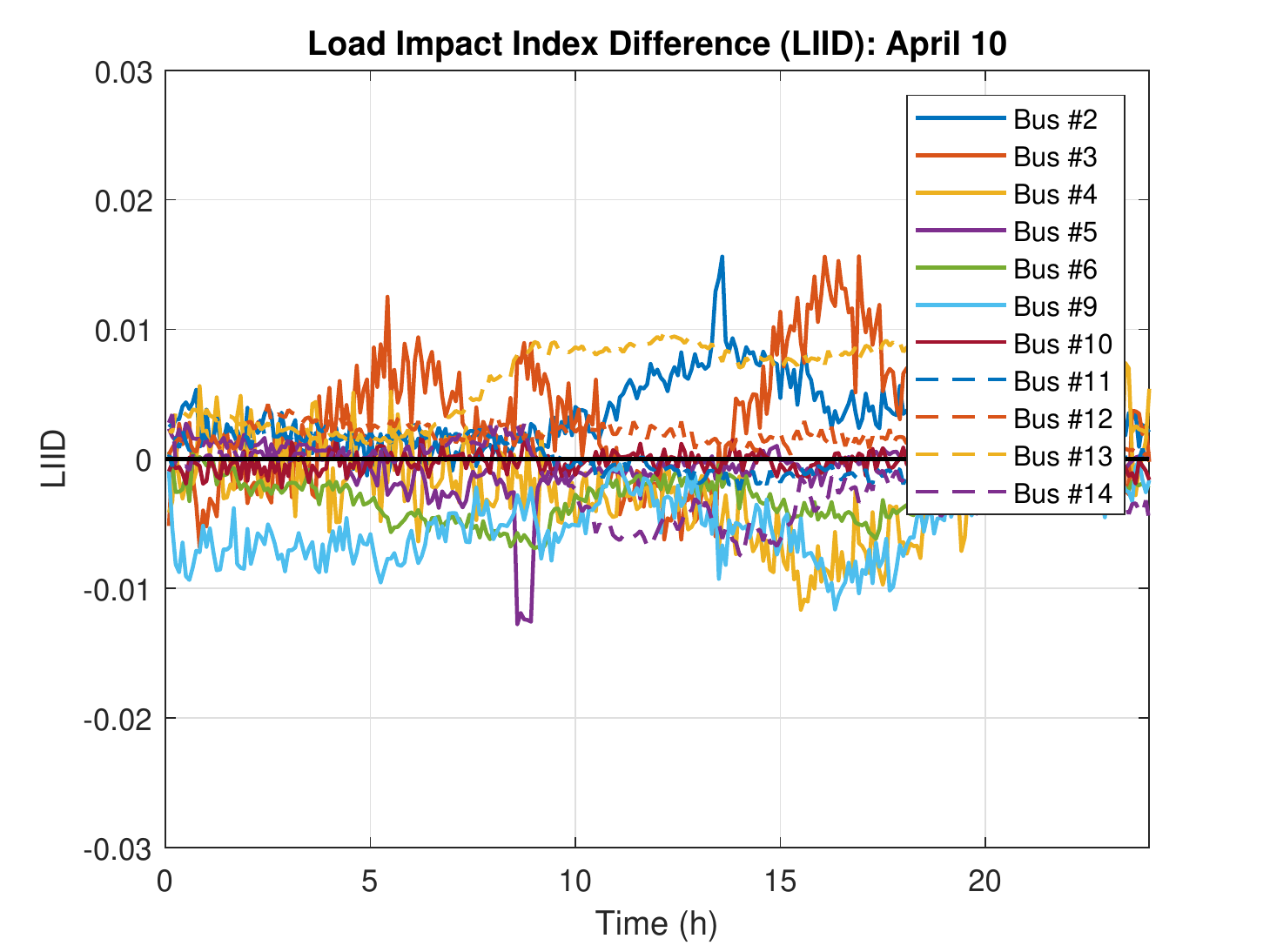}}
\\
\subfigure[] { \label{fig:liidapril11}     
\includegraphics[width=7.7cm]{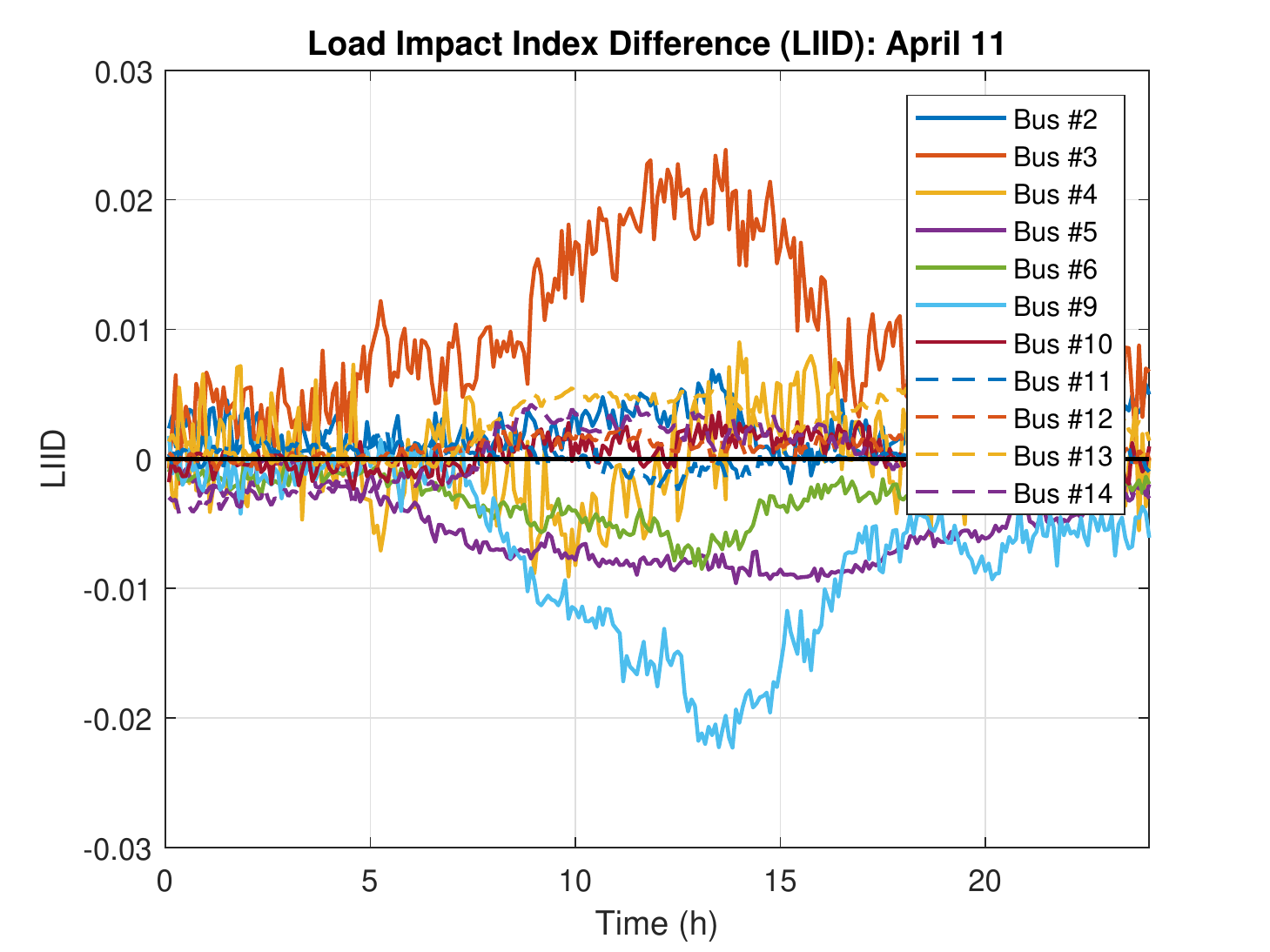}}
\subfigure[] { \label{fig:liidapril12}     
\includegraphics[width=7.7cm]{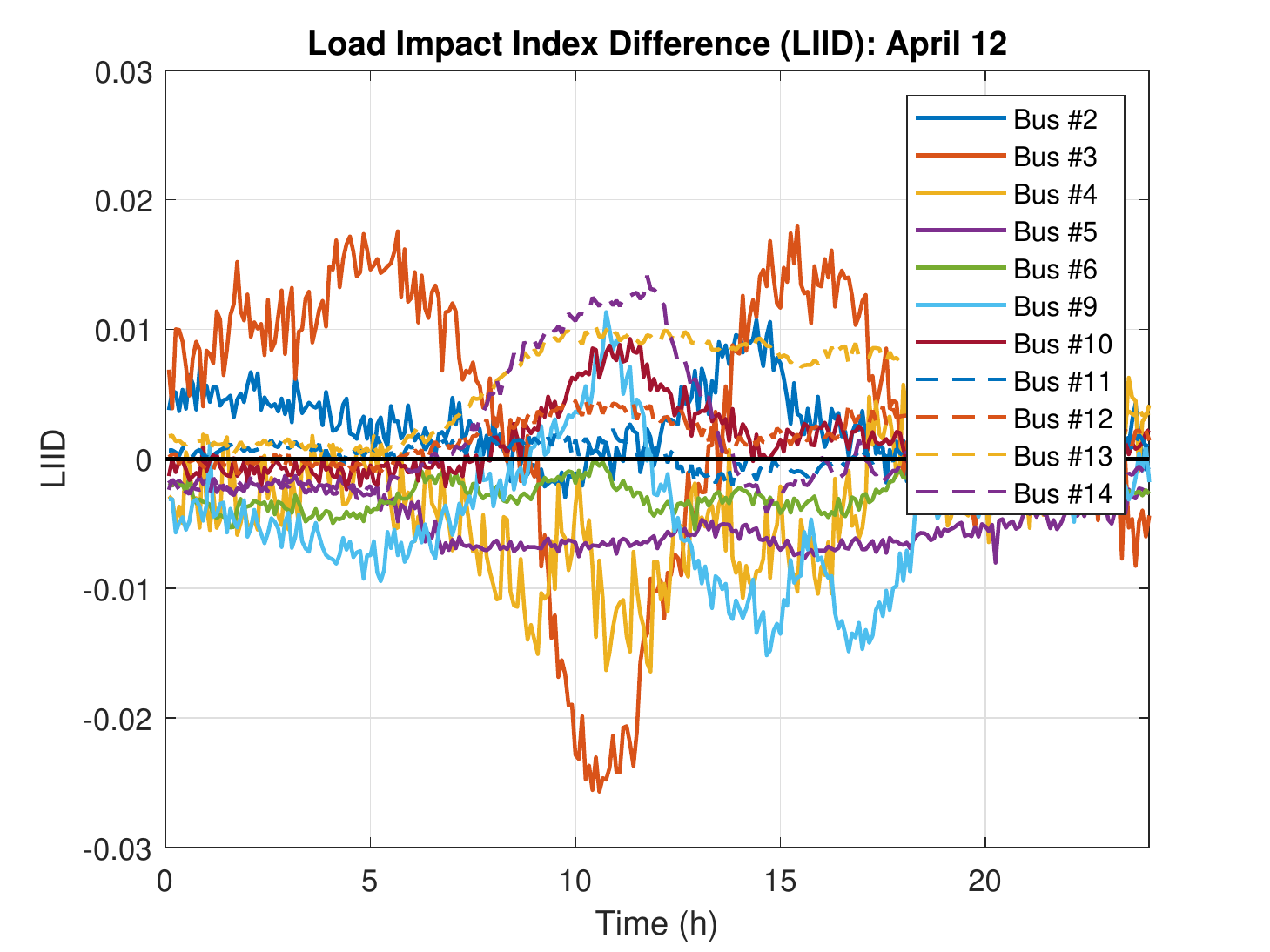}}
\caption{Load impact index difference (LIID) calculated for each bus in the system for the days: a) April 9, b) April 10, c) April 11, and d) April 12.} 
\label{fig:liidaprildays}  
\end{figure*}

\vspace{2mm}
\subsection{Case Study 1: April 9}

\subsubsection{Preliminary Analysis} 
A preliminary analysis is performed in order to identify the `ideal' time period and bus(es) that would need the minimum effort to cause a high impact in the system when a load-changing attack is performed; or simply the period when a peak of LD matches with a negative peak in LIID. Figs. \ref{fig:ldapril9} and \ref{fig:liidapril9} show the LD and the LIID for April 9. Based on the results from the analysis, the period when the highest difference in load demand between 2019 and 2020 arises is in the hours between 6:30 am to 7:30 am and 4:30 pm to 6:00 pm. The highest LD value during the morning period is 0.309. However, in order to find the discussed `ideal' period for the load-changing attack, we also need to find out the most negative LIID value observed during the examined periods and then correlate it to the high LD value period observed in the LD graph. As seen in Fig. \ref{fig:liidapril9}, the `lowest' negative LIID values occur during the 11:30 am to 1:30 pm period, with a peak of -0.0248. Nonetheless, when correlating these two values, we can clearly see that our LIID peak period does not match with any of the two LD identified periods, so no bus(es) in the system can be identified as the one(s) that could cause a high impact in the 2020 system when compared to the 2019 system scenario. Any load-changing attack implemented in 2020 system will have a similar effect on 2019 system, making April 9 a difficult day to analyze in terms of how more vulnerable the system is when low net-load demand conditions exist.

\subsubsection{Load-changing attack Impact}
According to the analysis obtained from the \textit{preliminary analysis} step, we recognized that for this particular day there are no time periods where an attacker could effectively compromise the frequency of the system when compared to 2019 with minimum effort. Both systems (2019 and 2020 scenarios) would have a fairly similar response, thus making it very hard to evaluate the feasibility of a load-changing attack when low net-loading conditions exist. This is indicated by the fact that the lowest (most negative) value of LIID does not align with the highest value of LD in the analyzed scenario. In a nutshell, April 9 is a day when an attacker would no see any significant differences between attacking a system with lower loading conditions, based on how the 2020 system scenario compares to the 2019 system scenario.

\subsection{Case Study 2: April 10}

\subsubsection{Preliminary Analysis} 
Figs. \ref{fig:ldapril10} and \ref{fig:liidapril10} show the LD and the LIID for April 10. Based on these results, we identify that the period when the highest difference in load demand between 2019 and 2020 occurs is in the hours between 5:00 am and 10:00 am, more specifically, at 7:30 am when the LD value is 0.356. Following a similar approach as the one presented in the previous case study, we correlate the most negative LIID value of the analyzed day, which occurs between 8:30 am and 9:00 am (i.e., the negative peak of the purple line), with the period of maximum LD. It is worthwhile to remember that the negative LIID value tells us which bus(es) in the system will have the greatest impact in the 2020 system while requiring the minimum effort (minimum load change required) when compared to the 2019 system. Based on the observed correlation, we conclude that attacking bus \#5 (purple line) between 8:30 am and 9:00 am (i.e., the negative peak of the purple line), together with bus \#9 (light blue line), would cause the greatest impact on the frequency stability of the 2020 system scenario.

\subsubsection{Load-changing attack Impact}
\begin{figure}[t]
\centerline{\includegraphics[width=0.85\linewidth]{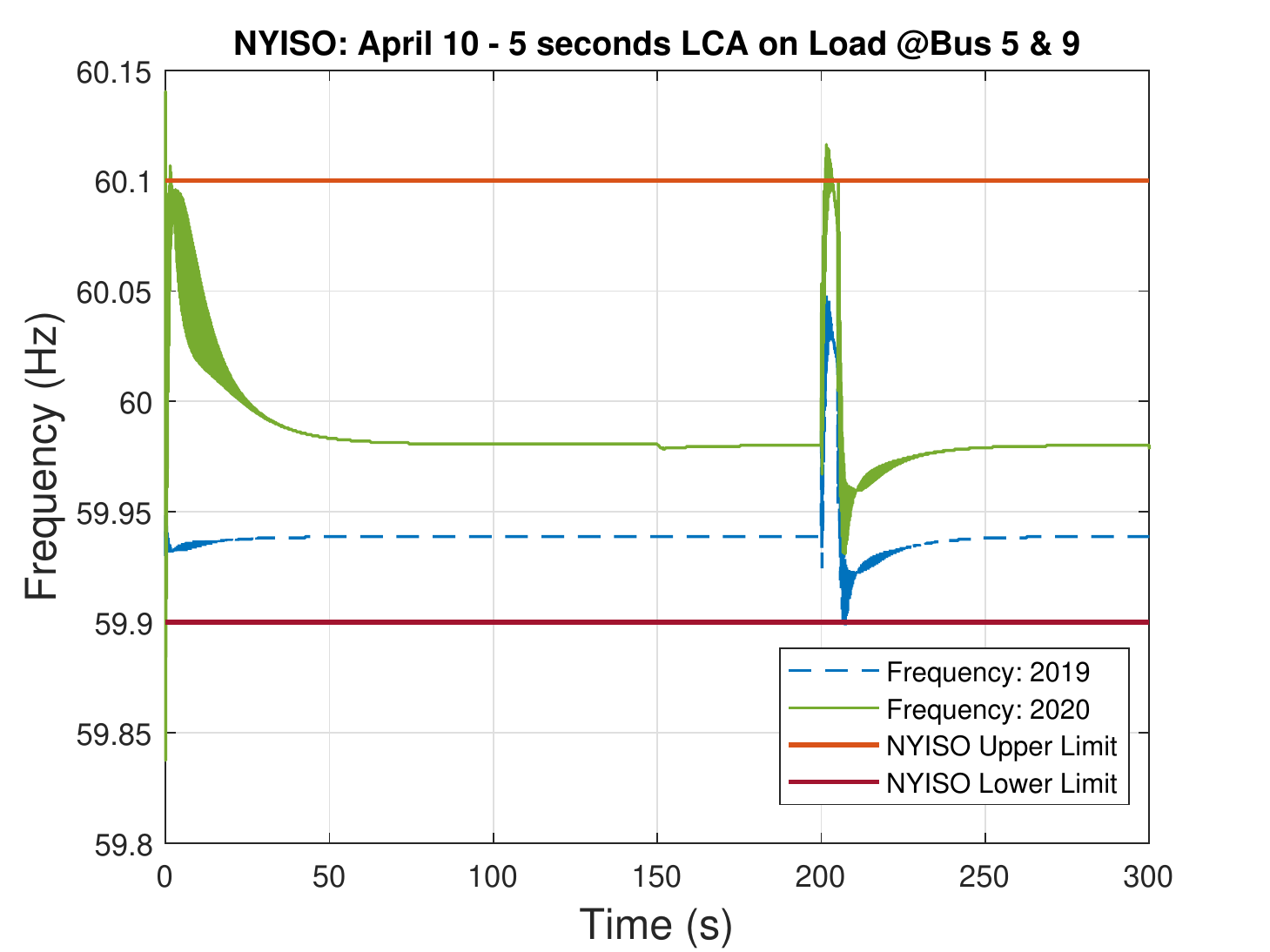}}
\caption{Load-changing attack impact on the frequency of the system during April 10 of 2019 and 2020. The upper and lower limits given by NYISO are provided for reference.} 
\label{fig:freqapril10}
\end{figure}

Based on the preliminary analysis, we simulate a load-changing attack on the loads connected to bus \#5 and bus \#9 for both the 2019 and 2020 system scenarios. Fig. \ref{fig:freqapril10} shows the impact of the 5-second load changing attack during 300 seconds (i.e., 5 minutes from 8:40 am to 8:45 am), where the load-changing attack is executed at 200 seconds. As seen in the graph, the frequency of the 2020 system scenario crosses the overfrequency NYISO limit of 60.1 Hz when the load-changing attack disconnects the compromised loads (i.e., loads at bus \#5 and bus \#9). On the other hand, the 2019 system scenario presents no overfrequency problems and thus we can see that the low net-load demand of the 2020 system makes more feasible the implementation of a high-impact load-changing attack that could compromise the stability of the system. It should also be noted that a more sustained attack (e.g., a 1 or 5-minute attack) has the potential of causing more severe problems.

\subsection{Case Study 3: April 11}

\subsubsection{Preliminary Analysis} 
Similar to April 10, April 11 seems to be a day where a load-changing attack could be feasible if mitigation measures are not taken. But, to confirm this, we perform the preliminary analysis and identify the highest LD values and the most negative LIID values based on Figs. \ref{fig:ldapril11} and \ref{fig:liidapril11}. These figures show the LD and the LIID for April 11. Based on these results, the period when the highest difference in load demand between 2019 and 2020 exists is in the hours between 7:30 am and 3:00 pm. Similarly to the previous case study, in order to find the `ideal' period for the load-changing attack, we need to find out the most negative LIID value observed during the examined period and then correlate it to the high LD value period observed in the LD graph. Correlating these two graphs, we observe that attacking bus \#9 (light blue line) between 1:00 pm and 2:00 pm (i.e., the negative peak of the light blue line) would cause the greatest impact on the frequency stability of the 2020 system scenario.

\subsubsection{Load-changing attack Impact}
\begin{figure}[t]
\centerline{\includegraphics[width=0.85\linewidth]{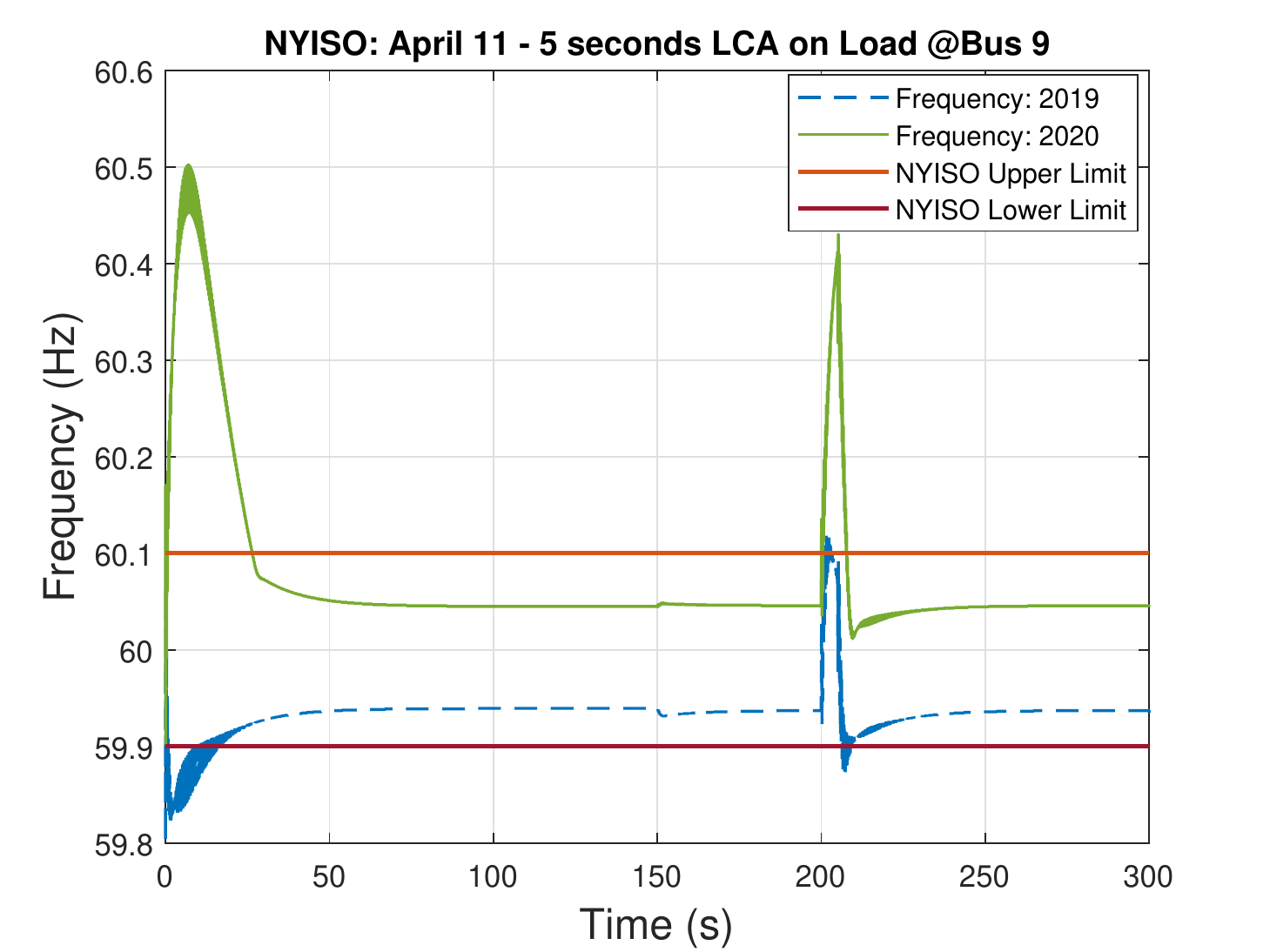}}
\caption{Load-changing attack impact on the frequency of the system during April 11 of 2019 and 2020. The upper and lower limits given by NYISO are provided for reference.} 
\label{fig:freqapril11}
\end{figure}

A load-changing attack on the load connected to bus \#9 is implemented for both the 2019 and 2020 system scenarios, according to the results obtained in the preliminary analysis. Fig. \ref{fig:freqapril11} shows the impact of the 5-second load changing attack during 300 seconds (i.e., 5 minutes from 1:40 pm to 1:45 pm), where the load-changing attack is executed at 200 seconds. This graph shows how the frequency of the 2020 system scenario crosses the overfrequency limit of 60.1 Hz and goes up to 60.5 Hz when the load-changing attack compromises the load connected at bus \#9. Differently, the 2019 system scenario presents no major overfrequency problems, so we can conclude again that the low net-load demand of the 2020 system makes more feasible the implementation of a high-impact load-changing attack that could negatively impact the stability of the system. 

\subsection{Case Study 4: April 12}

\subsubsection{Preliminary Analysis} 
Using a similar approach as the ones implemented in the previous case studies, a preliminary analysis is conducted to determine the `ideal' time period to attack the April 12 system scenario. Figs. \ref{fig:ldapril12} and \ref{fig:liidapril12} show LD and the corresponding LIID values for April 12. As seen in these graphs, the period when the highest difference in load demand between 2019 and 2020 appears is in the hours between 6:30 am to 12:30 pm, with a peak value of 0.628 at 8:10 am. This period can be correlated to the period when the most negative LIID values are observed, which in turn, is the period between 10:00 am and 11:30 am (i.e., the negative peak of the orange line with a peak negative value of -0.0257 at 10:35 am). Using this information, we conclude that attacking bus \#3 (orange line) between 10:00 am and 11:30 am would produce the greatest impact in the frequency stability of the 2020 system scenario, while requiring the minimum attacker's effort, when compared to the 2019 system.

\subsubsection{Load-changing attack Impact}
\begin{figure}[t]
\centerline{\includegraphics[width=0.85\linewidth]{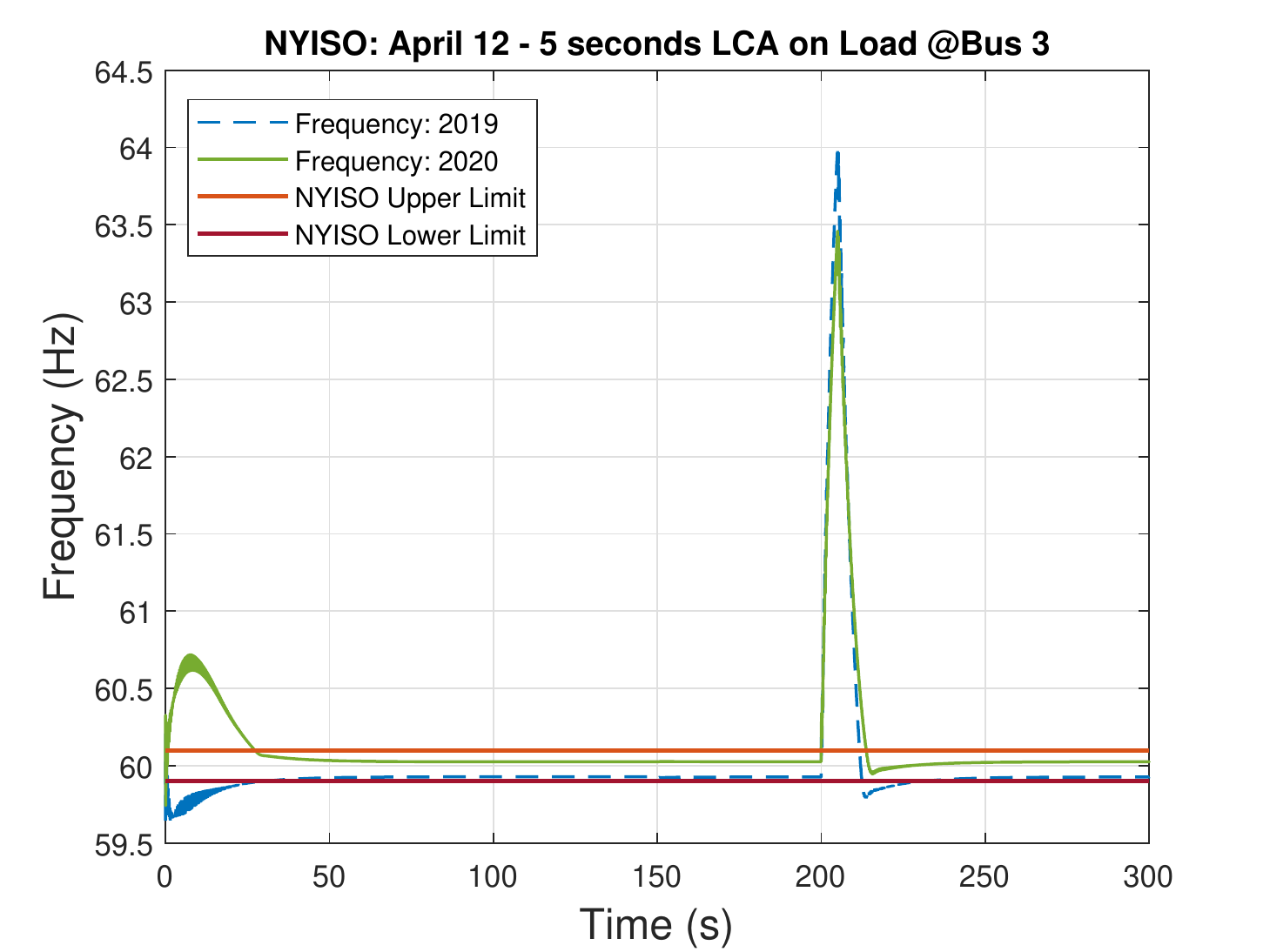}}
\caption{Load-changing attack impact on the frequency of the system during April 12 of 2019 and 2020. The upper and lower limits given by NYISO are provided for reference.} 
\label{fig:freqapril12}
\end{figure}

Following the same approach described in previous case studies and based on the preliminary analysis results, we perform a load-changing attack on the load connected to bus \#3 for both the 2019 and 2020 system scenarios. Fig. \ref{fig:freqapril12} shows the impact of the 5-second load changing attack during 300 seconds (i.e., 5 minutes from 10:33 am to 10:38 am), where the load-changing attack is executed at 200 seconds. Different from previous results, both the 2019 and 2020 systems have severe frequency stability problems since the frequency peaks at around 64.02 Hz and 63.46 Hz for the 2019 and the 2020 system scenarios, respectively. Diving deeper into these results, we discovered that bus \# 3 is one of the most critical buses in the system due to the fact that it represents an average of around 37\% of the total load of the 2019 system scenario and 36\% of the total load of the 2020 system scenario. This makes bus \#3 one of the most critical buses in the system, and the results that are shown in Fig. \ref{fig:freqapril12} clearly demonstrate that no matter if the system is experiencing low loading conditions (2020 system) or not (2019 system), a load-changing attack that disconnects this high-impact load zone would cause severe frequency fluctuations. In a nutshell, an attacker with enough capabilities to perform an attack targeted at bus \#3 of the analyzed system will cause severe problems in the frequency stability of the system.

\subsection{Discussion of Case Studies}
Based on the results of the  four case studies investigated, it can be observed that each one of them enables different insights regarding the feasibility of load-changing attacks in a CPES experiencing low loading conditions. Evaluating all these four cases provides a good perspective on some probable scenarios of the load-changing attack spectrum. 
For instance, case study 1 (April 9) demonstrates a scenario where an attacker, implementing a load-changing attack, would not perceive any difference between compromising a system experiencing low loading conditions, such as the 2020 scenario, and a nominal system, such as the 2019 scenario. On the other hand, both case study 2 and case study 3 (April 10 and April 11) clearly show how the low net-load demand conditions, that exist during the COVID-19 pandemic lockdowns, make the 2020 system scenario more vulnerable to low probability, high-impact load changing attacks when compared to the 2019 scenario. In both case studies, it can be observed how the same load-changing scenario has minor impacts on the 2019 system's frequency while producing major problems on the 2020 system's frequency. Finally, case study 4 (April 12) is an example of how compromising a critical bus in a CPES will cause severe frequency stability problems no matter if low loading conditions are being experienced. From a planning, operation, and cybersecurity perspective, system operators need to ensure these nodes are identified, and they must take appropriate mitigation strategies to handle any possible attacks to these critical buses.

\section{Conclusions and Future Work}\label{s:conclusion}

This paper explores the feasibility of load-changing attacks in CPES that experience abnormal low loading conditions caused by events such as the COVID-19 pandemic and its corresponding lockdown measures. We explore the differences in loading conditions of the main affected regions in the U.S. and analyze the abnormal load patterns caused by lockdown measures in these regions, with a primary focus on the NYSIO region, by applying DMD to load consumption data from the years 2019 and 2020. Based on these analyses, we formulate a load-changing attack and further explore the feasibility of such attack in a system experiencing the low loading patterns identified by the DMD process. Finally, we simulate and evaluate the impacts of load-changing attacks in a test grid system experiencing low loading conditions (2020), when compared to the 2019 historical loading conditions, using NYISO data. Our results demonstrate that low loading conditions can be leveraged by attackers with the objective of compromising the frequency stability of power systems. Specifically, the presented case studies show that an attacker with sufficient resources and capabilities would require less effort to compromise a system experiencing low loading conditions such as the ones experienced during the COVID-19 pandemic of 2020. 

Future work will focus on considering distribution systems with high penetration of renewable energy resources while evaluating both frequency and voltage stability problems caused by load-changing attacks. Systems with high penetration of intermittent renewable energy resources (such as PV and wind energy) are expected, due to the non-existent coordination of voltage regulation devices and the impacted system inertia (e.g, the removal of synchronous generators results in less system inertia with impacts on transient and small-signal stability), to be more vulnerable to load-changing attacks when considering their effects in other system areas and control routines such as voltage regulation. Another area of research will be the impact evaluation of other types of cyberattacks, such as false data injection attacks and time-delay attacks, in systems experiencing low loading conditions, such as the one investigated throughout this research.

\bibliographystyle{IEEEtran} 
\bibliography{biblio}
\vspace{6mm}
\section*{} \label{s:bio}

\vskip -2\baselineskip plus -1fil
\begin{IEEEbiography}[{\includegraphics[width=1in,height=1.25in,clip,keepaspectratio]{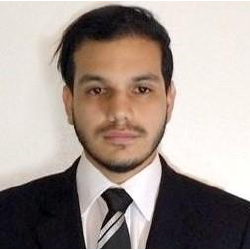}}]{Juan Ospina}~(S'13-M'20) is a Postdoctoral Research Associate with Florida State University and the Center for Advanced Power Systems, Tallahassee, Florida. He received a dual B.Sc. degree in Electrical and Computer Engineering in 2016, an M.S. in Electrical Engineering in 2018, and a Ph.D. in Electrical Engineering in 2019 from Florida State University, Tallahassee, Florida, USA. He is an IEEE and IEEE PES member. His research interests include the development of intelligent systems for electric power systems (EPS) and smart-grid applications, machine learning and reinforcement learning models for DER control, renewable energy integration, cybersecurity, and real-time simulation.
\end{IEEEbiography}

\vskip -2\baselineskip plus -1fil
\begin{IEEEbiography}[{\includegraphics[width=1in,height=1.25in,clip,keepaspectratio]{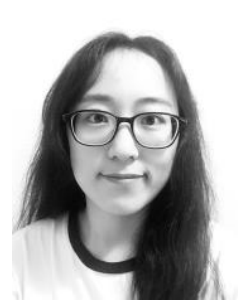}}]{XiaoRui Liu} (S'20) received her M.S. degree in Electrical Engineering from the Florida State University, Tallahassee, FL, USA, in 2017. She is currently pursuing the Ph.D. degree in Electrical Engineering at Florida State University, Tallahassee, FL, USA. Her research interest includes real-time simulation of power systems, cybersecurity, and machine learning. 
\end{IEEEbiography}

\vskip -2\baselineskip plus -1fil
\begin{IEEEbiography}[{\includegraphics[width=1in,height=1.25in,clip,keepaspectratio]{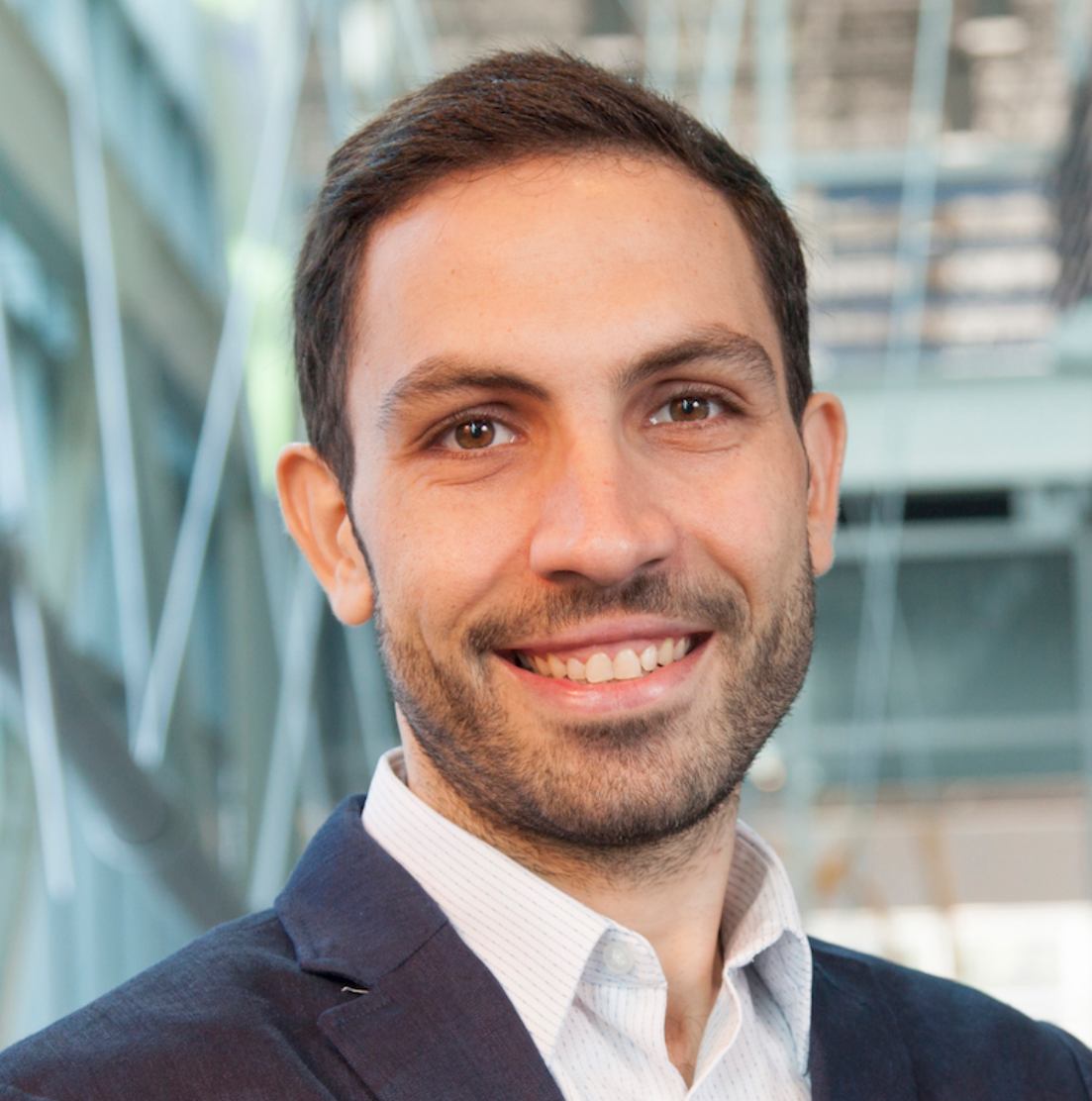}}]{Charalambos Konstantinou}~(S'11-M'18-SM'20) is an Assistant Professor of Electrical and Computer Engineering with Florida A\&M University and Florida State University (FAMU-FSU) College of Engineering and the Center for Advanced Power Systems, Florida State University, Tallahassee, FL. He received a Ph.D. in Electrical Engineering from New York University, NY, in 2018. His research interests include cyberphysical and embedded systems security with focus on power systems. He is the recipient of the 2020 Myron Zucker Student-Faculty Grant Award from IEEE Foundation, the Southeastern Center for Electrical Engineering Education (SCEEE) Young Faculty Development Award 2019, and the best paper award at the International Conference on Very Large Scale Integration (VLSI-SoC) 2018. 
\end{IEEEbiography}

\vskip -2\baselineskip plus -1fil
\begin{IEEEbiography}[{\includegraphics[width=1in,height=1.25in,clip,keepaspectratio]{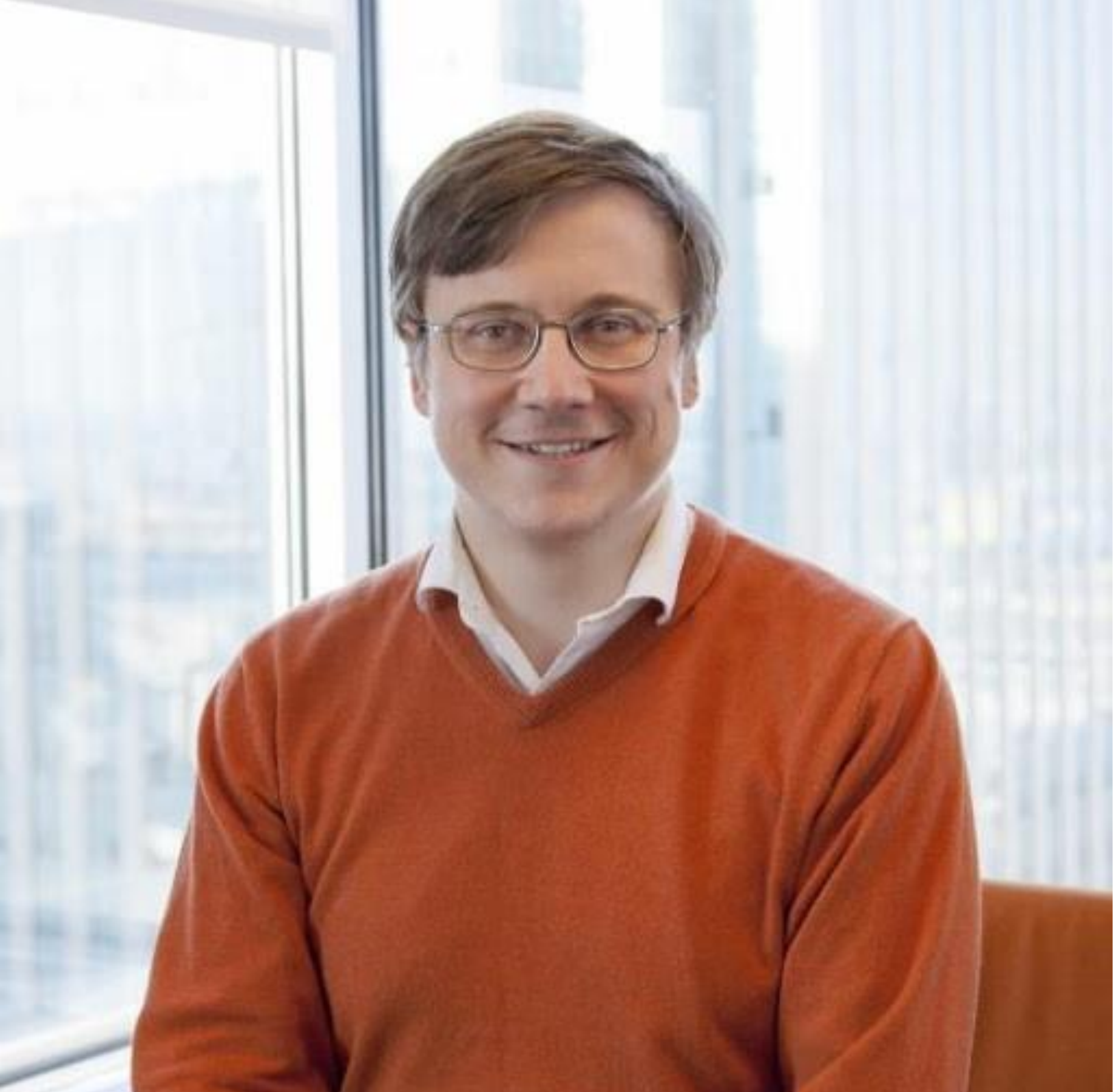}}]{Yury Dvorkin}~(S'11-M'20) is an Assistant Professor and Goddard Junior Faculty Fellow in the Department of Electrical and Computer Engineering at New York University’s Tandon School of Engineering and Center for Urban Science and Progress. His research work has been supported by the U.S. National Science Foundation, including the 2019 NSF CAREER Award, Alfred P. Sloan Foundation, U.S. Department of Energy and Advanced Research Projects Agency–Energy, and Electric Power Research Institute. Before joining NYU, he was a Ph.D. student at the University of Washington, under the supervision of Prof. Daniel S. Kirschen, and a graduate student researcher at the Center for Nonlinear Studies, Los Alamos National Laboratory, under the supervision of Dr. Michael Chertkov and Dr. Scott Backhaus. For his dissertation work, entitled “Operations and Planning in Sustainable Power Systems“, Yury was awarded the inaugural 2016 Scientific Achievement Award by Clean Energy Institute (University of Washington). Dvorkin’s research interests revolve around energy economics, civil infrastructure systems, and energy transition.
\end{IEEEbiography}

\EOD

\end{document}